\documentclass[preprint,3p,times,authoryear]{elsarticle}
\biboptions{numbers,sort&compress}
\usepackage[utf8]{inputenc}           % 如果用 XeLaTeX/LuaLaTeX 可删
\usepackage{amsmath,amssymb,amsfonts}
\usepackage{graphicx}
\usepackage{booktabs}
\usepackage{array}
\usepackage{enumitem}
\usepackage{bm}
\usepackage{siunitx}
\usepackage{xcolor}
\usepackage{url}
\usepackage{float}
\usepackage{hyperref}             
\usepackage{algorithm}
\usepackage{algpseudocode}% hyperref 放在 cleveref 之前或之后都可；常见先超链再 cleveref
\hypersetup{
  colorlinks=true,
  linkcolor=black,
  citecolor=black,
  urlcolor=black
}
\usepackage{cleveref}
\usepackage{lineno}                   % 行号宏包，实际开启放在 frontmatter 之后
\usepackage{subcaption} % 只保留 subcaption，不再加载 subfig
\usepackage{caption}
\usepackage{threeparttable}
\journal{IPM}

\crefname{equation}{Eq.}{Eqs.}
\Crefname{equation}{Eq.}{Eqs.}
\crefname{figure}{Fig.}{Figs.}
\Crefname{figure}{Fig.}{Figs.}
\crefname{table}{Table}{Tables}
\Crefname{table}{Table}{Tables}

\begin{document}

\begin{frontmatter}

\title{Exploring cooperation mechanisms via deep reinforcement learning in network common-pool resource games}
\author[aff1,aff2,aff3]{Yihang Qin}
\author[aff5]{Jinying Zou}
% \author[aff1,aff2,aff3,aff4]{Qi Su}
\author[aff1,aff2,aff3]{Lin Wang\corref{cor1}}
\cortext[cor1]{Corresponding author.}
\ead{wanglin@sjtu.edu.cn}

\address[aff1]{State Key Laboratory of Submarine Geoscience, Department of Automation, Shanghai Jiao Tong University, Shanghai
200240, China}
\address[aff2]{Key Laboratory for System Control and Information Processing, Ministry of Education of China, Shanghai 200240, China}
\address[aff3]{Shanghai Key Laboratory for Perception and Control in Industrial Network Systems, Shanghai 200240, China}
\address[aff5]{Kaiyuan International Mathematical Sciences Institute, Changsha 410008, China}

\begin{abstract}
Sustaining cooperation in networked common-pool resource systems requires allocation mechanisms that balance contribution incentives, resource sustainability, and distributional fairness. This paper proposes a network common-pool resource game in which each individual is embedded in overlapping local resource pools and faces endogenous resource constraints during strategy evolution. We first show that two typical allocation mechanisms lead to different failures: equal allocation drives cooperation and accumulated resources to nearly 0, whereas proportional allocation maintains only a cooperation level of about 0.5 and produces a high Gini coefficient of about 0.65. To search for adaptive allocation policies, we use a GNN-based deep reinforcement learning social planner as a mechanism-discovery tool rather than an end solution. 
In the 200-step evaluation on BA scale-free networks in Fig~\ref{fig:Agent_diagram}, the learned planner raises the final cooperation level to nearly 1 while maintaining a Gini coefficient of about 0.3.
We then interpret and distill the learned policy into two simpler mixture mechanisms: a resource-dependent mechanism for regular networks and a degree-conditioned mechanism for heterogeneous networks. The degree-conditioned mechanism achieves dynamics close to the learned planner, yields low inequality comparable to the full-cooperation equal-allocation benchmark in heterogeneous networks, and shows stronger robustness across network topologies and parameter shifts. 
These results suggest that resource-dependent and structure-aware allocation may provide useful principles for sustaining cooperation in networked resource-sharing  social systems.

% Cooperation is a fundamental organizing principle in biological and social systems. However, under resource constraints, cooperative behavior often collapses as defectors always gain resources unilaterally in interactions. To address this, we propose a Forced Loner Mechanism (FLM) integrated into the Spatial Prisoner’s Dilemma with resource dynamics (SPDL), where bankrupt agents are enforced to withdraw and receive a guaranteed subsidy. This mechanism models realistic social subsidies or industrial safety nets. Simulation results demonstrate that the forced Loner mechanism effectively restores cooperation when the temptation to defect is high, enhancing all the efficiency, sustainability, and fairness of the game system, which are evaluated through three macroscopic indicators: net output, average cumulative resources, and the Gini coefficient, respectively.  Moreover, a sensitivity analysis reveals that the mechanism is robust against variations in the Loner payoff. These findings provide a quantitative understanding of how social exit and protection mechanisms can be interventions to stabilize cooperation in resource-limited systems.
\end{abstract}


\end{frontmatter}

% \title{Recovering Cooperation in Spatial Prisoner’s Dilemma Game via Forced Loner}
% % \title{Forced Loner as a Robustness Mechanism in Spatial Prisoner’s Dilemma with Resources}
%  \author{
%     Yihang Qin, Hao Chen, and Lin Wang,~\IEEEmembership{Senior Member,~IEEE}%
%     \thanks{
%     % Manuscript received April 19, 2025; revised August 16, 2025. 
%     This work was supported in part by National Natural Science Foundation of China under Grants 62373245 and 12426311, and the National Key R\&D Program of China under Grant 2023YFB4706800, and in part by the “Dawn” Program of Shanghai Education Commission, China.}% <-this % stops a space
%     \thanks{Yihang Qin, Hao Chen and Lin Wang are with the School of Automation and Intelligent Sensing, Shanghai Jiao Tong University, Shanghai 200240, China. (e-mail:wanglin@sjtu.edu.cn)}%
%     % \thanks{Lin Wang(Corresponding author, e-mail: wanglin@sjtu.edu.cn) is with the School of Electronic Information, Shenzhen University, Shenzhen, China.}%
% }

 % \author{Yihang Qin, Lin Wang,~\IEEEmembership{Senior Member,~IEEE,}
        % <-this % stops a space


\section{Introduction}
Cooperation in human societies often depends on common resources that must be repeatedly allocated, used, replenished, and redistributed \cite{axelrod1981evolution,berg1995trust,neumann1944theory,nowak2006five,ohtsuki2006simple}. Governments allocate public funds to citizens and firms with the expectation of future taxes, innovation, or civic responsibility; organizations distribute resources to employees and teams whose effort sustains collective productivity; digital platforms provide visibility or shared infrastructure to users whose participation maintains the platform ecosystem; and ecological systems such as fisheries, forests, and irrigation networks require users to restrain short-term extraction to preserve future availability \cite{ross1973economic,grossman1986costs,chaudhuri2011sustaining,hardin1968tragedy,ostrom1990evolution,dietz2003struggle,holmstrom1982moral,fehr1999theory}. These situations share a common-pool resource problem: once resources are allocated, each recipient may either contribute back to the collective system or retain the benefits for private use. Although non-contribution can be individually beneficial in the short run, its spread gradually weakens resource regeneration, reduces future allocation capacity, and undermines long-term collective welfare \cite{ostrom2008challenge,agrawal2003sustainable,ostrom2002common,gardner1990nature}.

This problem becomes more challenging in networked populations. In many real systems, resources are not exchanged through a single global pool, but through local and overlapping interaction structures. Individuals may participate in different numbers of local pools, occupy unequal structural positions, and face heterogeneous risks of resource depletion. As a result, an allocation rule that appears fair or efficient within a single pool may generate unintended inequality or cooperation collapse at the population level. For example, equal allocation may provide local fairness but weaken incentives to contribute, whereas contribution-proportional allocation may encourage cooperation but amplify accumulated advantages. Therefore, the key challenge is not simply whether equality or proportionality is preferable, but how an allocation mechanism should adapt to both the current resource state and the structural position of local resource pools.
Recent studies have shown that reinforcement learning can be used to discover cooperation-promoting strategies and social mechanisms that are difficult to specify analytically. However, existing work still leaves an important question open for networked common-pool resource systems: can an adaptive allocation mechanism sustain cooperation when resource exchange is local, overlapping, and structurally heterogeneous? This question is different from designing individual strategies or changing network connections. Here, individual strategies continue evolving according to local payoff-based dynamics, while an external social planner only determines how the resources generated in each local pool should be redistributed. The planner must therefore influence long-term cooperation indirectly through resource allocation, rather than by directly controlling individual behaviour.
The objective of this study is to investigate how explainable deep reinforcement learning can be used to discover adaptive allocation mechanisms for networked common-pool resource systems. Specifically, this study addresses the following research questions:

\noindent\textbf{RQ1:} How do fairness-oriented equal distribution and contribution-based proportional mechanisms affect cooperation, resource sustainability, and inequality in networked common-pool resource systems with endogenous resource constraints?

\noindent\textbf{RQ2:} Can a learning-based social planner discover adaptive local allocation policies that improve cooperation and resource sustainability without directly controlling individual strategies?

\noindent\textbf{RQ3:} Can the learned allocation policy be interpreted and distilled into simpler mechanisms that reveal how redistribution should depend on local resource states and network structural positions?

To answer these questions, we formulate a network common-pool resource game in which each node is both an individual and a local resource pool. Individuals choose whether to cooperate or defect through evolutionary dynamics, while cooperators contribute resources to the pools centred on themselves and their neighbours. The grown resources in each pool are then allocated back to the corresponding local group. Importantly, each individual has an endogenous resource stock, and individuals with insufficient resources are unable to cooperate. This setting explicitly couples strategy evolution, resource accumulation, local interaction structure, and allocation mechanisms. The main contributions of this paper are as follows:
\begin{itemize}
\item A network common-pool resource game with endogenous resource constraints:
This paper develops a network common-pool resource game with overlapping ego-network resource pools and endogenous resource-constrained participation. The model captures how local allocation decisions affect not only immediate payoffs, but also individuals' future capacity to cooperate.

\item Learning-assisted mechanism discovery for resource allocation: 
We formulate local redistribution as a social-planning problem in which cooperation is influenced indirectly through resource allocation. A GNN-based RL planner is then used as a computational tool to search for adaptive allocation policies.

\item Interpretable mixture mechanisms distilled from the learned policy:
The learned policy is interpreted and distilled into two simpler mixture mechanisms: a resource-dependent mechanism for regular networks and a degree-conditioned mechanism for heterogeneous networks. These mechanisms show that effective redistribution should be both resource-dependent and structure-aware: peripheral nodes require protection from poverty traps, intermediate-degree nodes require contribution-based incentives, and hub-like nodes require redistribution to prevent excessive resource concentration.
\end{itemize}
 
The remainder of this paper is organized as follows. Section~2 reviews related work on cooperation mechanisms, learning-based mechanism design, and intelligent agents in networked systems. Section~3 introduces the network common-pool resource game and evaluation metrics. Section~4 presents the baseline allocation mechanisms and the proposed GNN-RL social planner. Section~5 reports the performance, interpretation, and mechanism distillation of the learned policy. Section~6 discusses the theoretical and practical implications, limitations, and future research directions. Section~7 concludes the paper.

\section{Related Work}

% A large body of research has investigated mechanisms that promote cooperation in social dilemmas and networked populations. Classical examples include voluntary participation \cite{szabo2002phase,hauert2002volunteering,hauert2002replicator,guo2020novel,jia2024freedom}, reputation and punishment \cite{nowak1998evolution,sigmund2001reward,szolnoki2011phase,szolnoki2017second,rand2011evolution,xia2023reputation}, prosocial exclusion \cite{sasaki2013evolution,li2015social}, network reciprocity \cite{nowak1992evolutionary,rand2011dynamic,su2022evolution}, and related institutional or structural mechanisms \cite{szolnoki2014cyclic,quan2021integrating}. These mechanisms can align individual incentives with collective welfare by changing outside options, social information, sanctions, or interaction patterns.

\subsection{Cooperation in social dilemmas and networked populations}

A large body of research has investigated mechanisms that promote cooperation in social dilemmas and networked populations. Classical examples include voluntary participation \cite{szabo2002phase,hauert2002volunteering,hauert2002replicator,guo2020novel,jia2024freedom}, reputation and punishment \cite{nowak1998evolution,sigmund2001reward,szolnoki2011phase,szolnoki2017second,rand2011evolution,xia2023reputation}, prosocial exclusion \cite{sasaki2013evolution,li2015social}, network reciprocity \cite{nowak1992evolutionary,rand2011dynamic,su2022evolution}, and related institutional or structural mechanisms \cite{szolnoki2014cyclic,hauser2014cooperating,hauser2019social,quan2021integrating}. These mechanisms can align individual incentives with collective welfare by changing outside options, social information, sanctions, participation opportunities, or interaction patterns. Recent studies have further shown that freedom of choice can disrupt cyclic dominance while maintaining cooperation in voluntary prisoner's dilemma games \cite{jia2024freedom}, and that opponent-specific responses can support the emergence of cooperation in repeated prisoner's dilemma interactions \cite{Morozov2026EmergenceOC}. These findings indicate that cooperation can be promoted not only by payoff incentives, but also by adaptive responses, social feedback, and institutional opportunities.

In networked populations, the structure of interactions plays a central role in shaping cooperation. Network reciprocity enables cooperators to form clusters and resist exploitation by defectors, while dynamic network adaptation allows individuals to modify their social connections in response to past interactions \cite{nowak1992evolutionary,rand2011dynamic,su2022evolution}. Beyond pairwise networks, recent studies have examined cooperation in higher-order and temporally evolving interaction structures. For example, \cite{Wang2026EvolutionOC} investigated public goods games on temporal higher-order networks and showed how historical utility and degree vectors can drive the evolution of interaction structures, which treat cooperation as a networked process shaped by structural information, individual attributes, and evolving interaction patterns. Recent work has also examined the tension between cooperation and fairness in networked public goods allocation \cite{sheng2026cooperation}. However, most existing mechanisms focus on changing strategic incentives, social information, punishment opportunities, participation options, or interaction structures. At the same time, less attention has been paid to how generated common resources should be redistributed to sustain future cooperation. This problem is especially important in networked common-pool resource systems, where resources are exchanged through local and overlapping pools and individuals occupy heterogeneous structural positions.

\subsection{Learning-based mechanism design and intelligent agents}

Recent advances in reinforcement learning have provided a computational route to explore cooperation-promoting strategies and mechanisms that are difficult to derive by hand. In evolutionary games, learning-based methods have been used to search large strategy spaces beyond classical heuristic rules, revealing adaptive strategies that can promote cooperation and improve social welfare in evolving populations \cite{sandholm1996multiagent,fatima2024learning,zheng2025evolution,xie2026reinforcement,lin2025coevolution,li2025memory,du2026shaping,jia2025asymmetric,su2025multi}. In particular, multi-agent reinforcement learning provides a flexible framework for studying how agents adapt their strategies in repeated and evolutionary games, and recent work has used this perspective to explore dominant strategies under different game environments \cite{su2025multi}. These studies show that reinforcement learning can serve not only as a model of adaptive decision making, but also as a computational tool for discovering behavioural rules that are difficult to identify through analytical reasoning alone.
More recently, this perspective has been extended from individual strategy learning to mechanism design, where a learning agent acts as an external social planner and searches for interventions that improve collective outcomes. For example, autonomous agents have been used to engineer social networks and increase cooperation in human groups \cite{shirado2020network}, deep reinforcement learning has been used to scaffold cooperation by recommending changes to social connections \cite{mckee2023scaffolding}, and dynamic resource-allocation mechanisms have been learned to promote sustainable reciprocation in common-pool resource problems \cite{koster2025deep}. Unlike \cite{koster2025deep}, who trained and evaluated a social planner using human behavioural data and behavioural clones in a small-group common-pool resource game, our study embeds the planner in a simulation-based evolutionary game on complex networks. Individual strategies evolve through the Fermi rule rather than human behaviors, while the planner operates over overlapping local pools and explicitly conditions allocation on network structure. These developments are part of the broader direction of deep mechanism design, in which deep reinforcement learning is used to discover social and economic policies that are difficult to specify analytically \cite{tacchetti2025deep}.
Related reinforcement-learning research has also considered fairness and operational constraints as explicit parts of the optimization problem. Multi-objective reinforcement learning can balance competing objectives such as efficiency and fairness through max--min or social-welfare formulations \cite{pmlr-v235-park24b,pmlr-v267-kim25ac}, while constrained reinforcement learning can incorporate feasibility, safety, or distributional requirements
directly into policy optimization. Recent work has further combined constrained Markov decision processes with graph neural networks for fairness-sensitive resource coordination in networked systems \cite{zhou2025federated}. These approaches provide complementary formulations when fairness or feasibility is specified explicitly as an optimization objective or constraint.

A related challenge is how learning systems represent and use network structure. Recent studies on graph-based learning have shown that GNNs can benefit from explicitly incorporating structural information, for example
through structure-aware augmentation and degree-aware aggregation \cite{he2026structure}. For decision-oriented AI systems, recent work has explored human-guided collective LLM intelligence, in which multiple specialized AI agents collaborate with retrieval-augmented information access to support strategic planning \cite{kim2026human, ChenExplainablePO}. Although these studies address different application domains, they reflect a common concern in the information processing community: AI systems should not only optimize task performance, but also process structural information, support collective decision-making, and provide interpretable or controllable decision processes \cite{Cambria2023ASO, Wang2024TheRO, Zhang2024TowardsEG}.

Nevertheless, existing learning-based cooperation studies still leave several questions open. Many studies focus on learning individual strategies, while social-planner studies often intervene by changing network connections or by designing allocation rules in small-group common-pool resource settings. Less attention has been paid to network-structured common-pool resource systems in which resources are generated and redistributed through overlapping local pools. Moreover, learned policies are often evaluated mainly by performance, whereas their underlying mechanism-design principles may remain difficult to interpret. The present study addresses this gap by training a graph neural network-based deep reinforcement learning social planner for local resource allocation, and by distilling the learned policy into interpretable resource-dependent and degree-conditioned allocation mechanisms.

\section{Game Model}

Understanding how cooperation emerges and is sustained in resource-constrained environments is a central problem in evolutionary game theory and socio-economic systems. Game Models such as the public goods game \cite{chaudhuri2011sustaining,szabo2002phase,szolnoki2011phase,sheng2026cooperation} and the common-pool resource game\cite{hardin1968tragedy,gardner1990nature,ostrom2002common,koster2025deep} have been proposed to describe and understand this problem. In many real-world scenarios, individuals are embedded in structured populations, where interactions are local, and resources are limited. Under such conditions, individuals’ strategic decisions not only influence their payoffs but also determine their ability to participate in future interactions. Furthermore, under what principles or mechanisms, whether fairness-oriented equal distribution, contribution-based proportional distribution, or some other approach, should resources be redistributed to foster and sustain cooperation?
To capture these coupled dynamics and problems, we propose a network common-pool resource game with endogenous resource constraints and allocation mechanisms. 
\begin{figure}[t]
    \centering
    \includegraphics[width=0.95\linewidth]{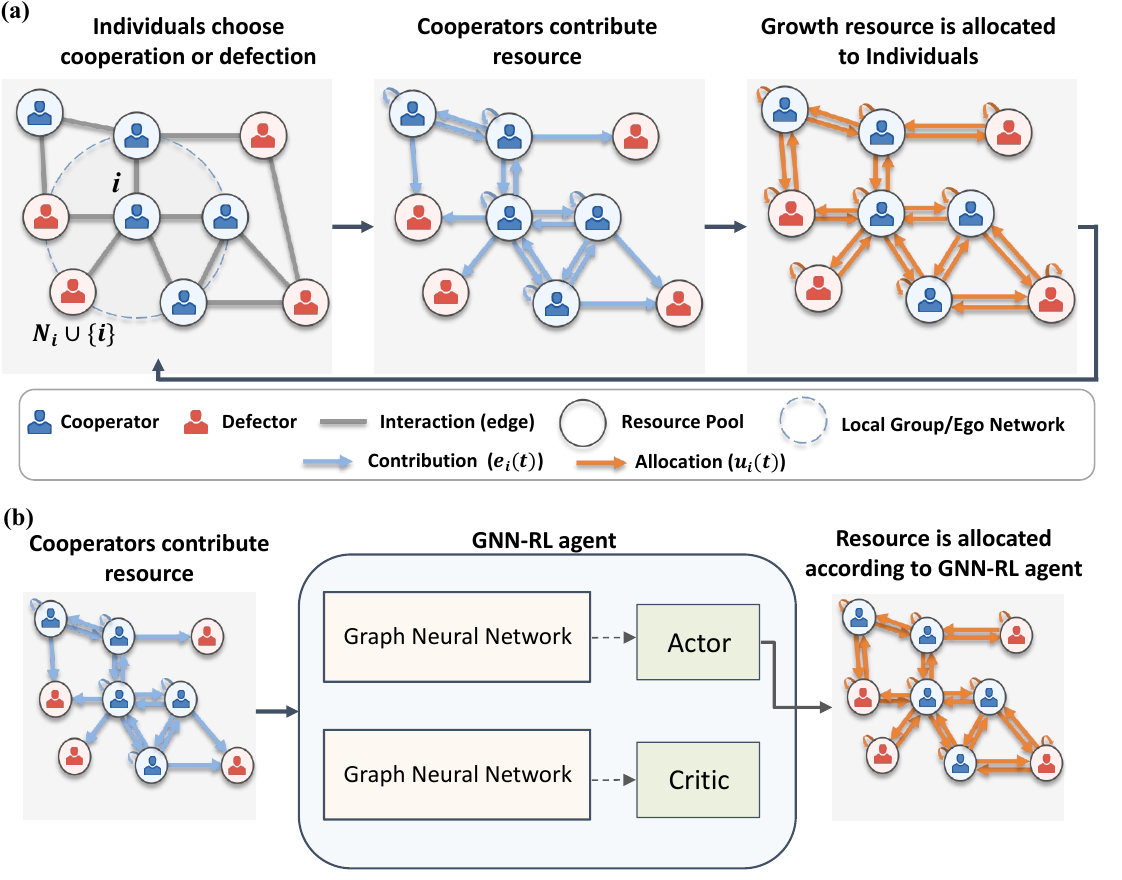}
    \caption{Overview of the network common-pool resource game.
    (a): In the network common-pool resource game, players are connected through a social network and decide whether to cooperate with or defect against their neighbours. Cooperators will contribute a certain amount of resources to the resource pools; the resources in the pools grow and are returned to players in their local group according to the allocation mechanism. (b): Our GNN-RL agent learns to act as a social planner, determining how to allocate resources for each resource pool in every round. It uses graph neural networks (GNN) to model the network common-pool game and employs reinforcement learning to optimize the allocation mechanism.}
    \label{fig:common-pool resource game}
\end{figure}
\subsection{Networked common-pool resource environment}
We consider a population of $N$ individuals, each with a resource pool structured by a social or economic network $G = (V, E)$,
where $V = \{1,2,\dots,N\}$ is the set of individuals and $E$ is the set of edges representing pairwise interactions. Each individual $i$ possesses resources $R_i(t)$, with an initial resource of $R_0$. At each time step $t$, individual $i$ selects a strategy $x_i(t) \in \{C,D\}$, where $C$ represents cooperation and $D$ represents defection.

If individual $i$ with degree of $d_i$ chooses cooperation, it will contribute a total resource of  
\begin{equation}
    e_i(t) = d_i+1 + \alpha\max\{0, R_i(t) - (d_i+1)\}, 
\end{equation}
uniformly to the resource pools of its ego network $N_i \cup \{i\}$, where $N_i=\{j\in V|(i,j)\in E\}$ denote the neighbor set of individual $i$. Each of the $d_i+1$ pools therefore receives $e_i(t)/(d_i +1)$,
where $\alpha$ is the resource contribution ratio. However, if an individual's resource level is below the minimum requirement \(d_i+1\), the individual cannot cooperate; we refer to this constraint as a poverty trap. If individual $i$ chooses defection, it will not contribute any resources, which means $e_i(t)=0$.

\subsection{Pool growth and baseline allocation mechanisms}
The resource pool associated with individual $i$ receives a total contribution of
\begin{equation}
    P_i(t)=\sum_{j \in N_i \cup \{i\}} \frac{e_j(t)}{d_j + 1}
\end{equation} from the members of its ego network accordingly and then grows at rate $r$ with maximum $P_{\max,i}(t)$. This reflects that resources, such as ecological stocks, public funds, or organizational resources, cannot expand without bound. The grown pool is
\begin{equation}
    G_i(t)=\min\{(1+r)P_i(t),P_{\max,i}(t)\},
\end{equation}
where $P_{\max,i}(t) = p_c \, N_{c,i}(t) \, \rho_{c,i}(t)$. Here, $p_c > 0$ is a scaling coefficient representing the basic productivity capacity per cooperator, $N_{c,i}(t)\geq0$ denotes the number of cooperators in pool $i$ at time $t$, and $\rho_{c,i}(t)\in[0,1]$ is the ego-network cooperation fraction to reflect the 
efficiency of cooperation defined as $\rho_{c,i}(t) = \frac{N_{c,i}(t)}{d_i + 1}$,

The resource $G_i(t)$ of the growth resource pool associated with individual $i$ is allocated to the members of its ego network according to certain allocation mechanisms. These are two classic allocation mechanisms within the socioeconomic system: fairness-oriented equal distribution and contribution-based proportional distribution.

Under the equal baseline, the allocable resources in each pool are uniformly distributed among all individuals in the corresponding ego networks. 
Specifically, individual $i$ receives $\frac{G_j(t)}{d_j+1}$ resource
from each resource pool  $j\in N_i\cup\{i\}$ of its ego network.
Therefore, the total resources received by individual $i$ from all the pools are given by
\begin{equation}
    u_i^{\mathrm{Equal}}(t)
    =
    \sum_{j\in N_i\cup\{i\}}
    \frac{G_j(t)}{d_j+1}.
\end{equation}

Under the proportional baseline, allocable resources in each pool are distributed according to the individuals' contributions within that ego network. Specifically, individual $i$ receives
$   \frac{e_i(t)/(d_i+1)}
    {\sum_{k\in N_j\cup\{j\}} e_k(t)/(d_k+1)}
    G_j(t)$
resource from each resource pool $j\in N_i\cup\{i\}$ of its ego network.
Accordingly, the total resources received by individual $i$ are
\begin{equation}
    u_i^{\mathrm{Prop}}(t)
    =
    \sum_{j\in N_i\cup\{i\}}
    \frac{e_i(t)/(d_i+1)}
    {\sum_{k\in N_j\cup\{j\}} e_k(t)/(d_k+1)}
    G_j(t).
\end{equation}

\subsection{Resource update and strategy evolution}
As mentioned before, individual $i$ contributes $e_i(t)$ to and receives resource $u_{i}(t)$ from a total of $d_i+1$ resource pools in its ego network.
In addition to resource inputs and outputs, each individual is subject to endogenous resource consumption, which captures the maintenance cost of sustaining local interactions and holding accumulated resources. It also motivates individuals to acquire resources continuously; otherwise, all individuals will maintain defection when they have sufficient resources that will neither decrease nor be consumed. 
Specifically, individual $i$ incurs a baseline consumption level proportional to its local interaction scale, together with an additional proportional consumption on the resource amount exceeding this baseline, $d_i$. The consumption function is defined as 
\begin{equation} 
c_i(t)=d_i+\tau \max\{0,\;R_i(t)-d_i\}, 
\end{equation} 
where $\tau \in [0,1]$ is the proportional consumption rate for the surplus resource beyond the baseline level $d_i$. The degree-dependent baseline term reflects that higher-degree individuals participate in more local interactions and therefore incur a larger maintenance burden. The linear form is adopted as a parsimonious modeling assumption rather than a universal empirical law.
Accordingly, the resource level of individual $i$ updates according to
\begin{equation}
R_i(t+1)=R_i(t)+u_i(t)-e_i(t)-c_i(t).
\end{equation}
During the evolutionary process, the minimum value of $R_i(t)$ is set to 0; if $R_i(t)$ is less than $d_i+1$, the individual cannot choose the cooperation strategy either.

Furthermore, the payoff of an individual $i$ is defined as the difference between income and input:
\begin{equation}
    r_{i}(t)=u_{i}(t)-e_{i}(t).
\end{equation}

The strategies evolve dynamically on the Fermi update rule according to the individuals' payoff. At each time step $t$, individual $i$ randomly selects a neighbour $j \in N_i$ and adopts its strategy with probability
\begin{equation}
P\big(x_i \leftarrow x_j\big) =
\frac{1}{1 + \exp\big((r_i(t) - r_j(t))/\kappa\big)},
\end{equation}
where $\kappa > 0$ controls the level of stochasticity. We illustrate the common-pool resource game in Fig.~\ref{fig:common-pool resource game}(a).

\begin{table}[!htbp]
\centering
\caption{Parameters used in the network common-pool resource game simulations.}
\label{tab:simulation_parameters}
\begin{tabular}{lll}
\toprule
Parameter & Description & Value(s) \\
\midrule
\(T\) & Simulation time steps & \(2\times10^4\) \\
\(N\) & Network size & \(50\) \\
\(\langle k\rangle\) & Average node degree & \(4\) \\
\(p_{\mathrm{ER}}\) & Edge probability of ER networks &  \(8.16 \times 10^{-2}\) \\
\(m\) & Attachment parameter of BA networks & \(2\) \\
\(k_{\mathrm{WS}}\) & Initial degree of WS networks & \(4\) \\
\(q_{\mathrm{WS}}\) & Rewiring probability of WS networks & \(0.1\) \\
\(R_0\) & Initial resource of each individual & \(20\) \\
\(r\) & Pool growth factor & \(0.5\) \\
\(p_c\) & Pool-capacity coefficient & \(50\) \\
\(\alpha\) & Resource contribution ratio & \(0.5\) \\
\(\tau\) & Resource consumption rate & \(0.05\) \\
\(\kappa\) & Fermi stochasticity parameter & \(1\) \\
\bottomrule
\end{tabular}
\end{table}
\subsection{Evaluation Metrics}
To evaluate the system's performance from multiple perspectives, we consider three key metrics: the cooperation level, the average accumulated resources, and the resource inequality.
First, the cooperation level is defined as the fraction of cooperators in the population at time $t$, denoted by $f_c(t)$, which reflects the overall level of cooperative behaviour in the game system.
Second, we measure the efficiency of the system by the average accumulated resources:
\[
\bar{R}(t) = \frac{1}{N} \sum_{i=1}^{N} R_i(t).
\]
This metric captures the overall productivity and sustainability of the game system.
Finally, we quantify the fairness of resource distribution using the Gini coefficient, defined as
\[
Gini = \frac{1}{2N^2 \bar{R}(t)} \sum_{i=1}^{N} \sum_{j=1}^{N} |R_i(t) - R_j(t)|,
\]
which measures the inequality of resource allocation among individuals. A lower value of $Gini$ indicates a more equitable distribution. 
In numerical implementation, the denominator is replaced by \(2N^2(\bar{R}(t)+\epsilon)\) with a small \(\epsilon>0\).
Together, these three metrics provide a comprehensive evaluation of cooperation, efficiency, and fairness in the  system.

% Unless otherwise specified, we set the evolutionary dynamics parameters as follows: $N=50$, $r=0.5$,$T=2\times10^4$, $R_0=20$, $p_c=50$, $\alpha=0.5$, $\tau=0.05$, and $\kappa=1$.

\section{From baseline allocation mechanisms to learning social planner}

\subsection{Baseline mechanisms and preliminary observations}
\begin{figure}[t]
    \centering
    \includegraphics[width=0.95\linewidth]{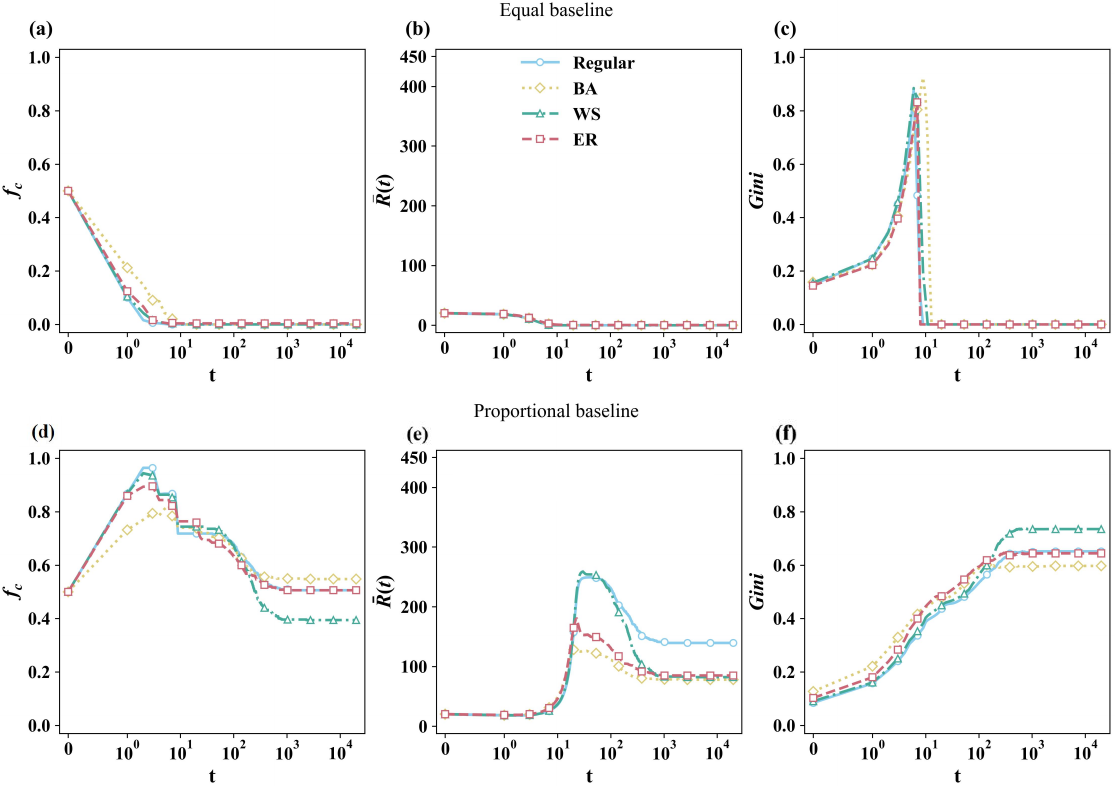}
    \caption{Evaluation metrics over time under different network topologies. (a)-(c) are under the equal baseline. (d)-(f) are under the proportional baseline.  Curves are averaged over 50 independent evaluation runs.   The equal baseline undermines initiative and encourages free-riding, thereby weakening cooperation; the proportional baseline, on the other hand, can promote cooperation to a certain extent, but it would trap the defector in a poverty trap, making it impossible for them to cooperate ever again. }
    \label{fig:Equal_Prop_3}
\end{figure}
\begin{figure}[t]
    \centering
    \includegraphics[width=0.95\linewidth]{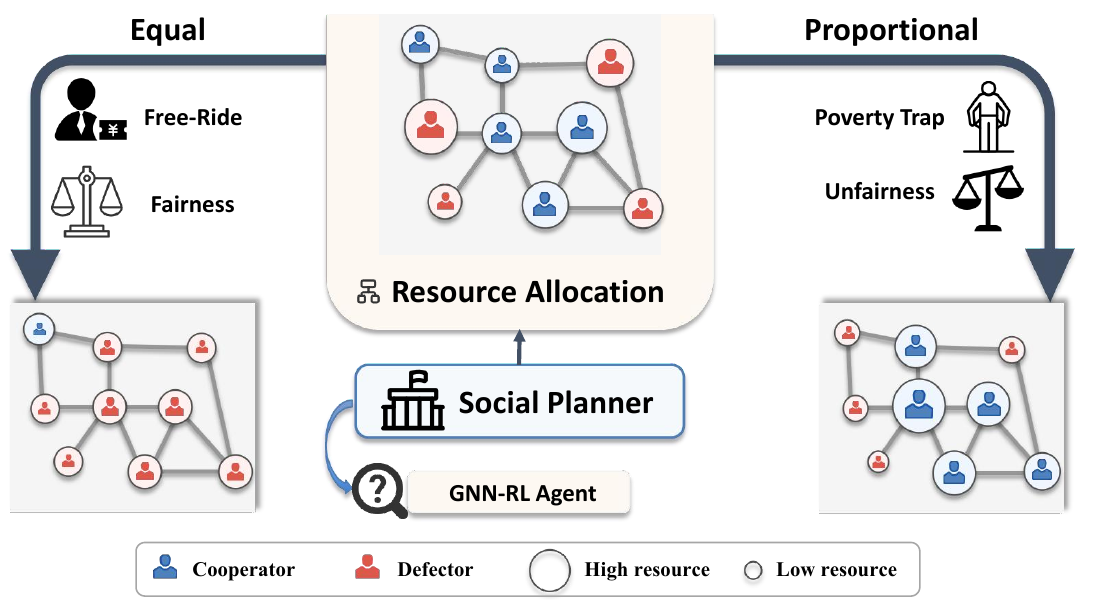}
    \caption{Illustration of the resource-allocation problem and the limitations of two baseline mechanisms. The social planner allocates resources generated in local pools to individuals in the network. Under equal allocation, resources are distributed fairly across individuals, but weak contribution incentives may drive the population into cooperation collapse. Under proportional allocation, contribution incentives are strengthened, but those with fewer resources can no longer participate in the cooperation, which may lead to Matthew effects and severe inequality. Node colors denote strategies (blue: cooperator; red: defector), and node sizes denote accumulated resource levels.} 
    \label{fig:reference_points}
\end{figure}

Before introducing the proposed reinforcement learning framework in Fig.~\ref{fig:common-pool resource game}(b), we first consider representative baseline allocation mechanisms to provide reference points for evaluating the effectiveness of learned strategies.
Fig.~\ref{fig:reference_points} conceptually illustrates the role of the social planner and the contrasting consequences of equal and proportional allocation.
Specifically, we examine the equal baseline in Fig.~\ref{fig:Equal_Prop_3}(a)-(c) and the proportional baseline in Fig.~\ref{fig:Equal_Prop_3}(d)-(f).
The baselines are selected at the mechanism level, as our objective is to compare alternative resource-redistribution principles rather than learning optimizers. Alternative RL algorithms would primarily test optimization sensitivity rather than constitute different allocation mechanisms. Accordingly, we do not conduct an exhaustive comparison across RL algorithms, as such benchmarking would primarily address optimizer selection rather than the mechanism-design question studied here.

We evaluate both mechanisms under four different network topologies, including regular networks, Erdős–Rényi (ER) random networks, Barabási–Albert (BA) scale-free networks, and Watts–Strogatz (WS) small-world networks. Unless otherwise specified, we set the evolutionary dynamics parameters as shown in Table~\ref{tab:simulation_parameters}. To ensure a fair comparison across different network topologies, the average node degree is fixed at $\langle k \rangle = 4$ in all cases. Moreover, initial strategies are assigned randomly with \(f_C(0)=f_D(0)=0.5\). The results are obtained by averaging 50 independent simulations.

Under the equal allocation mechanism, resources are distributed uniformly, regardless of individual contributions. 
Equal allocation is locally egalitarian because each pool distributes its available resources uniformly among all members, irrespective of their contributions. However, equal allocation at the pool level does not necessarily guarantee equality in the accumulated resources of the entire network after the evolutionary dynamics unfold.
Equal allocation impedes individual incentives for contribution. Individuals often tend to free-ride. 
Consequently, individuals gradually lose motivation to cooperate, resulting in a rapid decline in cooperation levels in Fig.~\ref{fig:Equal_Prop_3}(a). 
Ultimately, the system collapses into a state dominated by defection, where cooperative behaviour is no longer sustained. As shown in Fig.~\ref{fig:Equal_Prop_3}(b)(c), the average accumulated resources $\bar{R}(t)$ have fallen to near $0$, meaning everyone is equally poor. In the Regular, WS, and BA networks, the Gini and the $f_c$ approach $0$, indicating that almost no one cooperates and everyone is equally poor; in the ER network, the Gini coefficient is close to 1, but $f_c$ and $\bar{R(t)}$ remain close to $0$, meaning that only a very small number of individuals possess resources, while everyone else has none.
This observation highlights a fundamental limitation of purely fairness-oriented allocation: while equality is achieved, the lack of incentive mechanisms undermines the sustainability of cooperation.
By contrast, the proportional allocation mechanism rewards individuals based on their contributions, thereby providing a direct incentive for cooperation. 
Initially, this mechanism promotes cooperative behaviour, leading to a rapid increase in the level of cooperation in Fig.~\ref{fig:Equal_Prop_3}(d) from $t=0$ to $t=10$.
However, if an individual chooses to defect, its contribution becomes zero, and it consequently receives no resource under proportional allocation. Once its resource level falls below the cooperation threshold $d_i+1$, it will never again have the opportunity to cooperate, and its resources will remain at zero, while cooperators can continue to invest resources and grow increasingly wealthy.
At the population level, it further generates a Matthew effect in resource accumulation: trapped defectors remain at very low resource levels, whereas persistent cooperators continue to receive resources, contribute, and accumulate wealth.

As more defectors enter this absorbing poverty trap, the resulting separation between defectors and cooperators produces a Matthew effect, which is reflected in the steadily increasing Gini coefficient in Fig.~\ref{fig:Equal_Prop_3}(f).

Furthermore, despite maintaining a certain level of cooperation, the system's overall performance remains suboptimal. 
The average accumulated resource $\bar{R}(t)$ first increases and then decreases in Fig.~\ref{fig:Equal_Prop_3}(e), and does not reach its theoretical benchmark (see~\ref{Appendix:Analysis of Steady State}), indicating inefficiencies in resource utilization.
These results suggest that, while contribution-based allocation can sustain cooperation to some extent, it can also lead to unequal resource distribution and inefficient resource utilization.
Since neither the equal baseline nor the proportional baseline can maintain a high level of cooperation while ensuring a relatively fair allocation of resources, we employ a reinforcement learning approach to learn a new allocation mechanism as shown in Fig.~\ref{fig:common-pool resource game}(b).
\begin{figure}[!htbp]
    \centering
    \includegraphics[width=0.95\linewidth]{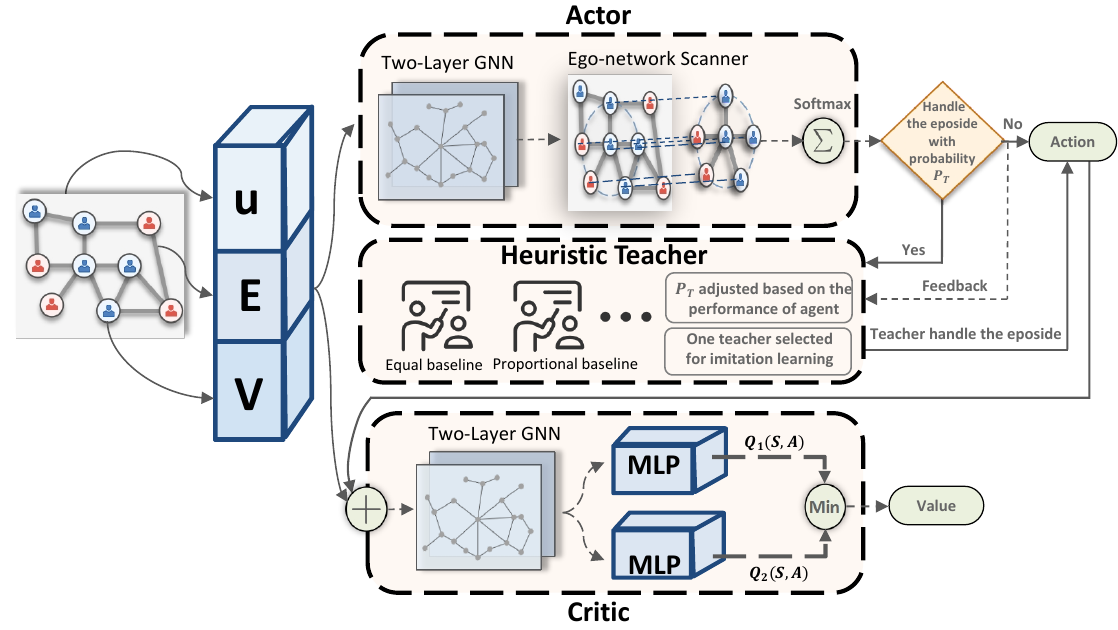}
    \caption{GNN-RL Agent architecture.
    The agent observes global features $u$, node features $V$, and edge features $E$ from the game state. 
    The actor first encodes the graph with a two-layer GraphNet backbone, then applies a shared ego-network allocation head to each focal node, functioning like a convolution kernel.
    Due to the randomness of evolutionary games and their sensitivity to allocation strategies, we introduce heuristic teachers' strategies to accelerate and stabilize the learning process.}
    \label{fig:GNN-RL}
\end{figure}
\subsection{Reinforcement learning framework}
To learn an adaptive resource allocation mechanism, we use a graph neural network-based reinforcement learning (GNN-RL) social planner as a computational tool for mechanism discovery, as illustrated in Fig.~\ref{fig:GNN-RL}. Different from traditional evolutionary game models in which the allocation mechanism is manually specified, the proposed framework acts as a social planner, learning an allocation mechanism from repeated interactions with the network common-pool resource environment.
The GNN-RL agent does not directly determine whether individuals cooperate or defect. Instead, it acts as a social planner. At each time step, every individual first chooses their strategies according to the evolutionary game process described in Section~3. Cooperators contribute resources to local pools, and the GNN-RL agent then determines how the grown pool resources should be allocated back to the individuals in each ego network, as shown in Fig.~\ref{fig:common-pool resource game}(b).
The framework consists of three main modules: an observation graph module, an actor network, and a critic network.

At each time step, the current game system state is represented as the environment state, and
we represent the current game state as a graph tensor representation $(u, V, E)$.
The global feature vector $u \in \mathbb{R}^{4}$ contains graph-level summary statistics, including the fraction of cooperators $f_c$, the mean accumulated resource $\bar{R}(t)$, the mean resource pool value $\bar{G}(t)$, and the Gini coefficient.
The node feature matrix $V \in \mathbb{R}^{N \times 4}$ contains, for each node $i$, its resource pool $G_i(t)$, accumulated resource $R_i(t)$, degree $d_i$, and the contribution resource $e_i(t)$.
The edge feature tensor $E \in \mathbb{R}^{N \times N \times 1}$ records the connectivity information used by the model. Specifically, each entry indicates whether node $j$ is included in the feasible ego-network of node $i$.

The actor network is responsible for generating allocation actions. It first encodes the global structural and state information of the network using a two-layer GraphNet backbone. Then, for each focal node $i$, which defines a local resource pool over its ego-network, a shared ego-network allocation head is applied to the subgraph consisting of node $i$ and its neighbours. Based on the learned local representations, the actor produces allocation scores for all feasible participants in the pool, and a softmax operation normalizes these scores into allocation weights. Repeating this process for all focal nodes yields the final allocation action over all local pools.
This design matches the local interaction structure of the common-pool resource game. Since each resource pool is defined within a local neighbourhood, the ego-network allocation head enables pool-level decision making while sharing parameters across the entire graph.

The critic network evaluates the quality of the allocation actions generated by the actor. It takes the current graph state and the allocation action as input, and estimates the expected long-term return. To improve training stability and reduce overestimation bias, we adopt a twin-critic structure. Two critic networks independently estimate the action value, and the smaller value is used during policy evaluation. This design follows the common practice in actor--critic algorithms for continuous control problems.
\subsection{Training design}
% \begin{table}[t]
% \centering
% \caption{Main training parameters of the proposed GNN-RL planner.}
% \label{tab:main_training_parameters}
% \begin{tabular}{ll}
% \toprule
% Training parameter & Value \\
% \midrule
% Total training environment steps & $5 \times 10^7$ \\
% Episode length $T$ & 200 \\
% Batch size & 256 \\
% Actor learning rate & $5\times10^{-6}$ \\
% Critic learning rate & $2\times10^{-5}$ \\
% Discount factor $\gamma$ & 0.99 \\
% Target update rate $\rho$ & 0.005 \\
% Policy delay & 8 \\
% Gradient steps per update & 2 \\
% Reward signal & $\bar{R}(t+1)$ \\
% \bottomrule
% \end{tabular}
% \end{table}
Training the social planner in the proposed evolutionary game is inherently challenging because the environment is highly non-stationary. Individuals continuously update their strategies during an episode, and the planner's allocation decisions influence not only the immediate state but also the subsequent evolutionary trajectory of the system. In particular, poor allocation decisions in the early stage of an episode may drive the population into a low-cooperation regime that is difficult to recover from later. To limit this long-horizon instability and improve optimization tractability, we use relatively short episodes of 200 steps. 

To improve training stability and expose the planner to structurally diverse environments, we train the planner over multiple graph topologies rather than a single fixed interaction structure. Specifically, the training distribution comprises four topology families, including regular networks, ER networks, BA networks, and WS networks. For each topology, we construct a fixed graph bank using predetermined random seeds. During training, graph instances are sampled from these fixed banks. This design increases structural diversity while avoiding the excessive variance that would arise if entirely new graphs were regenerated at every reset. As a result, the planner is exposed to heterogeneous network structures while the optimization process remains reproducible and more stable.
To further reduce the instability caused by uninformative exploration,
we adopt a staged teacher-guided training procedure. The teachers consist of an equal baseline, a proportional baseline, and a pool-dependent interpolation baseline that adaptively combines uniform and
contribution-proportional allocation according to the current pool state \cite{koster2025deep}. The complete definition of this rule is provided in \ref{Appendix:Interpolation baseline}.
During online TD3 training, teacher-guided actions are initially retained to
prevent early cooperation collapse. Teacher control is then released
gradually after the actor demonstrates sufficiently stable performance. At
the same time, the imitation-learning constraint is progressively weakened and
the actor's value-maximization objective is strengthened. This staged
handoff allows the planner to benefit from a stable initial allocation
mechanism while eventually learning an autonomous policy that can improve
beyond the heuristic teacher.
The actor architecture is further designed to match the local interaction pattern of the common-pool resource game. After global message passing over the entire graph, the final policy head is implemented as a shared ego-network allocation head, which is applied repeatedly to each focal node and its local neighbourhood. In this way, each focal node defines a local resource pool, and the actor produces a row-wise allocation distribution over the corresponding local participant set. This convolution-inspired design introduces a strong locality prior: allocation decisions are made at the pool level, while the same local decision mechanism is shared across all nodes in the graph. Besides improving generalization across graph instances, this design also makes the learned policy easier to interpret, since the final action can be understood as repeated applications of a common local allocation mechanism. In other words, a social planner applies the same allocation mechanism to the ego network of every node in the entire graph.
For the optimization objective, we use a resource-oriented planner reward. 
At each step, the planner receives a scaled reward proportional to the population-mean resource level in the next state, denoted by \(\lambda_R\bar{R}(t+1)\).
The training objective is to maximize the cumulative return over the episode. This reward design is motivated by the fact that sustained resource accumulation depends on continued cooperative replenishment through the network dynamics. Therefore, although the reward is defined in terms of system-wide resources, it also serves as an indirect objective for sustaining cooperation over time. We adopt the standard TD3 update scheme, including delayed policy updates and soft target-network updates. The pseudocode for the algorithm model is shown below Algorithm~\ref{alg:gnn_rl_teacher}.
The detailed network architecture and training hyperparameters are provided in~\ref{Appendix:Network Architecture Specification}. 
% The main training hyperparameters are reported separately in 
% Table~\ref{tab:main_training_parameters}. For a detailed network architecture, see~\ref{Appendix:Network Architecture Specification}.

\begin{algorithm}[t]
\caption{GNN-RL Agent learning with heuristic teacher intervention in networked common-pool resource games}
\label{alg:gnn_rl_teacher}
\begin{algorithmic}[1]
\Require Game parameters, fixed graph bank $\mathcal{B}$, maximum episodes $M$,
episode length $T_E$, teacher intervention probability $P_T$;
Actor $\pi_{\theta}$, twin critics $Q_{\psi_1},Q_{\psi_2}$, heuristic teacher set $\Pi_T$;
\Ensure Trained allocation policy $\pi_{\theta}$;

\State Initialize actor $\pi_{\theta}$, critics $Q_{\psi_1},Q_{\psi_2}$, target networks, and replay buffer $\mathcal{D}$;
\State Initialize teacher intervention probability $P_T$;

\For{episode $m=1$ to $M$}
    \State Sample a graph $G$ from the fixed graph bank $\mathcal{B}$;
    \State Initialize individual strategies and resources on $G$;
    \State Generate $\xi \sim U(0,1)$;
    \If{$\xi < P_T$}
        \State Select a heuristic teacher $\pi_T \in \Pi_T$ to handle this episode;
    \Else
        \State Use the actor policy $\pi_{\theta}$ to handle this episode;
    \EndIf

    \For{$t=1$ to $T$}
        \State Individuals update strategies according to the evolutionary game rule;
        \State Cooperators contribute resources to local pools;
        \State Construct the current graph state $s_t=(u_t,V_t,E_t)$;

        \If{the episode is handled by the teacher}
            \State Generate allocation matrix $A_t \leftarrow \pi_T(s_t)$;
            \State Mark $A_t$ as a teacher action;
        \Else
            \State Generate allocation matrix $A_t \leftarrow \pi_{\theta}(s_t)$;
        \EndIf

        \State Allocate grown pool resources according to $A_t$;
        \State Observe reward $r_t$, next state $s_{t+1}$, and terminal flag $d_t$;
        \State Compute planner reward
            \[
                r_t=\lambda_R \overline{R}(t+1),
            \]
        \State Store transition $(s_t,A_t,r_t,s_{t+1},d_t)$ in $\mathcal{D}$;

        \State Sample a minibatch from $\mathcal{D}$;
        \State Update twin critics using the clipped double-Q TD target;
        \If{delayed policy update step}
            \State Update actor using critic feedback and teacher imitation regularization;
            \State Soft-update target networks;
        \EndIf
    \EndFor

    \State Evaluate the actor policy $\pi_{\theta}$;
    \State Adjust teacher intervention probability $P_T$ and  regularization coefficient according to actor performance;
\EndFor

\State \Return trained allocation policy $\pi_{\theta}$.
\end{algorithmic}
\end{algorithm}

\subsection{Ablation of teacher-guided training}
\begin{table}[t]
\centering
\caption{Ablation of teacher-guided training. 
E, P, and I denote the Equal, Proportional, and Interpolation teachers, respectively. A training run was regarded as successful if the evaluation cooperation level exceeded $f_c$>0.9. $f_C$ is reported as mean $\pm$ standard deviation over all 8 independent training runs}
\label{tab:teacher_ablation}
\begin{tabular}{lcccc}
\toprule
Training setting & Teacher set & Imitation & Success rate & $f_C$ \\
\midrule
No Teacher
& None
& No
& $0/8$
& $0.052 \pm 0.037$
 \\

Full Teacher
& E + P + I
& No
& $0/8$
& $0.244 \pm 0.092$
\\
Equal + Proportional
& E + P
& Yes
& $0/8$
& $0.611 \pm 0.124$
 \\

Full Teacher
& E + P + I
& Yes
& $7/8$
& $0.944 \pm 0.038$
 \\
\bottomrule
\end{tabular}
\end{table}
To examine the role of teacher guidance during policy learning, we conducted an ablation study by progressively removing components of the teacher set. Specifically, we compared unguided training, training with only the Equal and Proportional teachers, and the full teacher setting that additionally includes the Interpolation teacher. All other network settings, model architectures, and training hyperparameters were kept unchanged. For teacher-guided training, the imitation coefficient was gradually annealed to zero during the first $50 \%$ of training and remained zero throughout the latter half. These results suggest that the Interpolation teacher primarily serves as an early-stage training scaffold that helps the RL agent escape the low-cooperation regime. Importantly, the imitation regularization is completely removed during the latter half of training, such that the final-stage actor updates are no longer directly constrained toward the teacher policy.

Overall, the proposed GNN-RL framework learns an adaptive allocation mechanism over structured populations. By combining graph neural networks with reinforcement learning, the model can exploit local topological information while optimizing long-term collective outcomes.

\section{Performance and interpretation of the RL agent's policy}
\begin{figure}[t]
    \centering
    \includegraphics[width=0.95\linewidth]{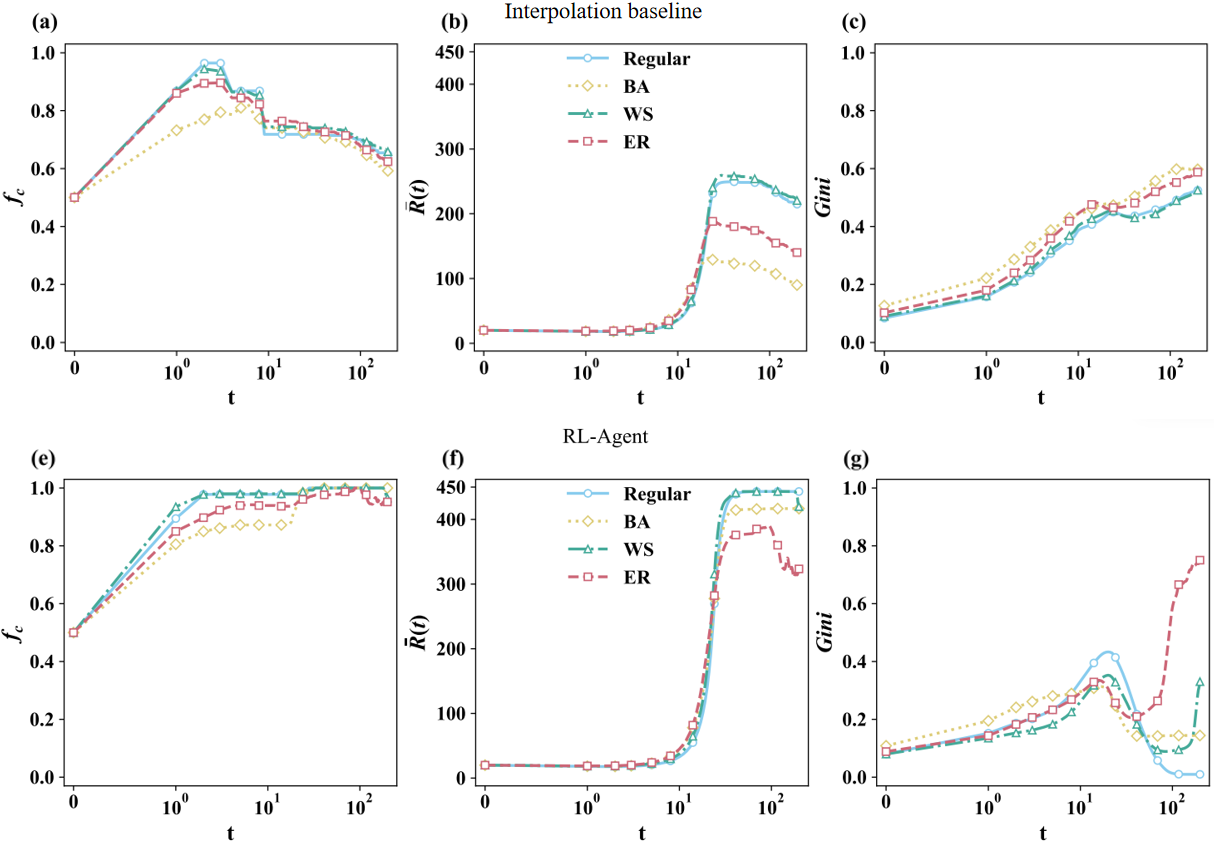}
    \caption{Evaluation metrics over time under different network topologies for the Interpolation and RL-agent. Here, we used the same topology seeds and the episode length ($200$) for the Interpolation baseline. Curves are averaged over 50 independent evaluation runs. The RL agent can learn a good allocation strategy to significantly enhance cooperation, increase accumulated resources, and reduce the Gini index, despite some fluctuations at the end of the round on certain network types.}
    \label{fig:Agent_diagram}
\end{figure}
\subsection{Performance of the learned GNN-RL social planner}
We first evaluate the performance of the learned GNN-RL social planner in the training environment. The episode length is 200, and the topology seeds are the same as those used during training. Fig.~\ref{fig:Agent_diagram} shows the temporal evolution of the cooperation level $f_c(t)$, the average accumulated resource $\bar{R}(t)$, and the Gini coefficient under different topologies. In each panel, different curves correspond to different network topologies, including Regular, ER, BA, and WS networks.
For the WS and ER networks, $\bar{R}(t)$ and the Gini coefficient exhibit visible fluctuations during the later stage of the episode, especially for $t=100$--$200$. This is because the curves are averaged over multiple independent evaluation runs, and a small number of runs under these two topologies show relatively poor performance in the later stage.
Nevertheless, compared with the equal baseline, the GNN-RL planner maintains a much higher cooperation level and prevents the system from rapidly collapsing into a low-resource state. This suggests that the learned planner preserves stronger incentives for cooperation than purely uniform redistribution. Compared with the proportional baseline, the GNN-RL planner achieves a more balanced outcome. It sustains a higher level of accumulated resources while maintaining a lower Gini coefficient.

Overall, the learned GNN-RL social planner improves cooperation, resource sustainability, and fairness across different network topologies. These results indicate that the trained planner does not simply reproduce either equal or proportional allocation. Instead, it learns an adaptive allocation policy that can balance cooperation incentives and resource redistribution according to the evolving system state.

\subsection{Interpretation of the learned allocation mechanism}
\begin{figure}[t]
    \centering
    \includegraphics[width=0.95\linewidth]{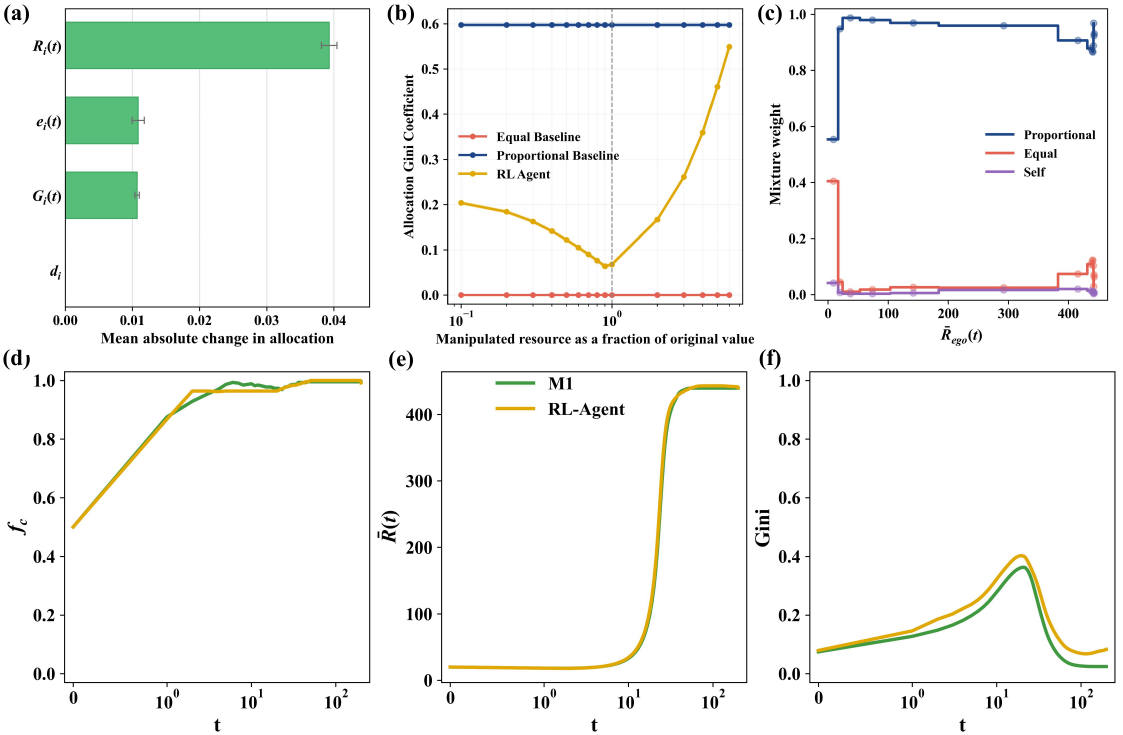}
    \caption{Interpretation of the learned allocation mechanism on regular networks.
    (a) Counterfactual feature-importance analysis of the actor's allocation output. Error bars denote SEM (Standard Error of the Mean) over 1,000 state snapshots.
    (b) Single-variable intervention on accumulated resource, where the resource feature is manipulated as a fraction of its original value.
    For each sampled pool-state observation, the focal node's perceived resource is multiplied by a factor $m$, with $m=1$ denoting the original observation. Resource-related observation channels are updated consistently, while the actual investment, pool state, cooperation state, and network structure are held fixed. The vertical axis shows the within-pool Gini coefficient of the resulting allocation proportions. Each line denote means over 2,048 sampled pool-state observations from each allocation mechanism episode, and shaded regions indicate $\pm1$ SEM.
    (c) Fitted mixture weights of the resource-binned mixture mechanism as a function of the ego-network average accumulated resource $\bar{R}_{\mathrm{ego}}(t)$. Solid step lines denote the fitted mixture weights, and circular markers denote the corresponding empirical estimates/data points.
    (d)--(f) Comparison between the original RL-Agent and the interpretable mixture mechanism M1 in terms of cooperation level, average accumulated resource, and Gini coefficient. Here, we used the same topology seeds and the length of episode ($200$) as during training. Results are averaged over 50 independent evaluation runs.}
    \label{fig:regular_diagram}
\end{figure}

After evaluating the performance of the learned GNN-RL social planner, we further investigate what allocation mechanism it has learned. We first focus on regular networks, where all nodes have the same degree. This setting removes the influence of degree heterogeneity and allows us to identify the state variables that directly affect the learned allocation policy.

We begin with a counterfactual feature-importance analysis. For each input feature in the ego-network representation, we set its value to zero while keeping the remaining features unchanged, and then measure the resulting change in the actor's allocation output. As shown in Fig.~\ref{fig:regular_diagram}(a), the accumulated resource $R_i(t)$ has the strongest influence on the learned allocation decisions. In contrast, the contribution resource $e_i(t)$, the grown pool value $G_i(t)$, and the degree $d_i$ have weaker effects. The weak effect of degree is expected in regular networks, since all nodes have the same structural role.
To further verify the role of accumulated resource, we conduct a single-variable intervention by manipulating the accumulated resource feature as a fraction of its original value. In this intervention, only the resource feature observed by the actor is modified. The actual contribution $e_i(t)$ and the environment dynamics are still computed from the original resource level before intervention. Fig.~\ref{fig:regular_diagram}(b) shows that the allocation behaviour of the RL agent changes systematically under a one-step counterfactual intervention on the focal node's perceived resource level, whereas the equal and proportional baselines remain fixed because their allocation rules do not depend on the manipulated resource-related observations (state samples are drawn from 2,048 samples per mechanism). The within-pool allocation Gini changes markedly across intervention levels, indicating that the learned actor adjusts the concentration of its allocation according to the perceived resource-related state. This provides evidence that the learned planner actively uses resource-related information when making allocation decisions within ego networks.

Based on the above observation, we construct an interpretable resource-binned mixture mechanism, denoted by M1. Since the actor applies a shared ego-network allocation head to scanning every focal node, we interpret the learned policy at the ego-network level. For the resource pool centered at node $i$, we define the mean accumulated resource of its ego-network as
\[
\bar{R}_{\mathrm{ego},i}(t)
=
\frac{1}{d_i+1}
\sum_{j\in N_i\cup\{i\}} R_j(t).
\]
This quantity summarizes the local resource state of all individuals participating in pool $i$. We then discretized $\bar{R}_{\mathrm{ego},i}(t)$ using quantile-based binning, which results in 14 effective resource bins, and approximate the actor's allocation behaviour within each bin as a mixture of three representative allocation mechanisms: equal allocation, proportional allocation, and self-allocation.

For the resource pool centered at node $i$, let $\hat{A}_{i\to j}(t)$ denote the allocation proportion assigned to individual $j\in N_i\cup\{i\}$. Therefore, the actual amount of resources allocated from pool $i$ to individual $j$ is $\hat{A}_{i\to j}(t)G_i(t)$. The allocation proportion of M1 is expressed as
\[
\hat{A}_{i\to j}^{\mathrm{M1}}(t)
=
\omega_{\mathrm{Equal}}^{(b_i(t))}
\frac{1}{d_i+1}
+
\omega_{\mathrm{Prop}}^{(b_i(t))}
\frac{e_j(t)/(d_j+1)}
{\sum_{k\in N_i\cup\{i\}} e_k(t)/(d_k+1)}
+
\omega_{\mathrm{Self}}^{(b_i(t))}
\mathbf{1}_{j=i},
\]
where $b_i(t)$ denotes the bin index determined by $\bar{R}_{\mathrm{ego},i}(t)$. For each bin, the three mixture weights are constrained to be non-negative and sum to one. The three coefficients 
$\omega_{\mathrm{Equal}}^{(b_i(t))}$, 
$\omega_{\mathrm{Prop}}^{(b_i(t))}$, and 
$\omega_{\mathrm{Self}}^{(b_i(t))}$ represent the relative weights of equal allocation, proportional allocation, and self-allocation in the corresponding resource bin, respectively. To obtain the interpretable mixture weights, we perform a post-hoc behavioural decomposition of the frozen RL policy. For each ego-network state associated with the resource pool centered at node i, the allocation vector produced by the RL actor is approximated as a convex combination of proportional allocation, equal allocation, and self-allocation using least-squares fitting that minimizes the squared $L_2$ reconstruction error. The fitted coefficients are subsequently projected onto the probability simplex to ensure non-negativity and unit sum. The mean absolute fitting error is $1.95 \times 10^{-3}$ per individual in the ego-network, indicating that the three-component representation closely approximates the allocation behaviour of the RL actor. More details are provided in the Appendix.

Fig.~\ref{fig:regular_diagram}(c) shows the fitted mixture weights under different ego-network resource bins defined by $\bar{R}_{\mathrm{ego},i}(t)$. The learned policy is not equivalent to a purely equal or purely proportional mechanism. Instead, we interpret it as a resource-dependent mixture of several simple allocation principles. In regular networks, the proportional component remains dominant over most ego-network resource bins, while the equal and self-allocation components provide additional state-dependent adjustments. The proportional component dominates over most resource regimes. In the low-resource regime, the increase in the Equal component is well identified and indicates stronger protective redistribution. In several high-resource bins, however, the Equal and Proportional basis vectors become increasingly similar, as reflected by the larger spectral condition numbers reported in~\ref{Appendix:M1_fitting}.

Finally, we evaluate whether this interpretable mixture mechanism can reproduce the performance of the original RL agent. As shown in Fig.~\ref{fig:regular_diagram}(d)--(f), M1 produces dynamics close to those of the RL-Agent in terms of cooperation level, average accumulated resource, and Gini coefficient. Although M1 is much simpler than the neural policy, it preserves the main performance advantage of the learned planner. 
This suggests that, on regular networks, the main behaviour of the GNN-RL planner can be captured by a resource-dependent mixture mechanism.
The resource-conditioned nature of M1 is conceptually related to the Interpolation mechanism of Koster et al. (2025). This similarity is expected in the regular-network setting, where structural heterogeneity is absent. We therefore regard M1 primarily as an intermediate resource-dependent representation of the learned policy; the additional mechanism-design insight emerges in heterogeneous networks, where the policy becomes explicitly degree-conditioned, as captured by M2.
% 加上度的影响因素以后是怎么样的，以Scale-Free为例来理解。
\begin{figure}[t]
    \centering
    \includegraphics[width=0.95\linewidth]{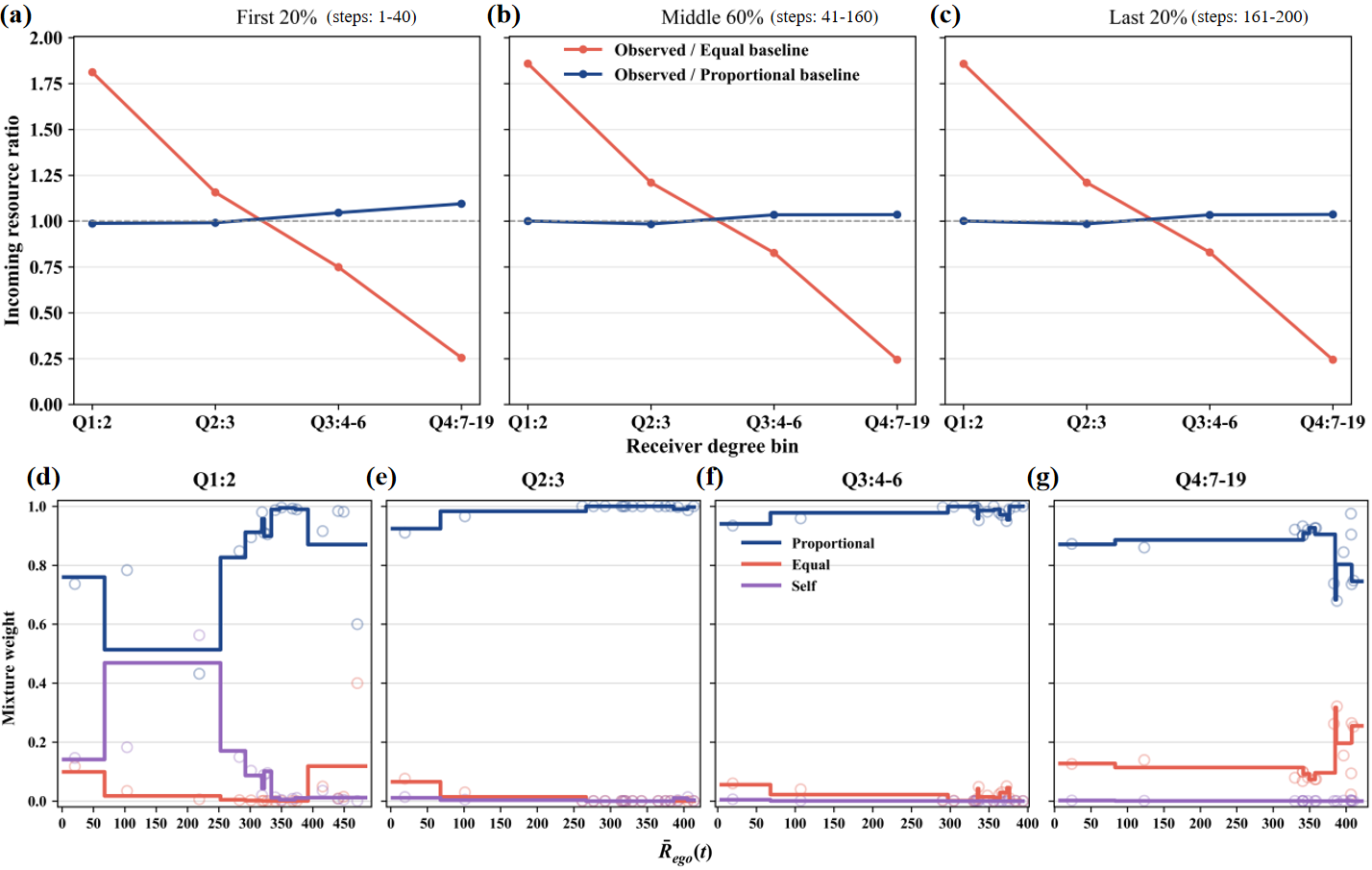}
    \caption{Interpretation of the learned allocation mechanism on scale-free networks. (a)-(c): degree-dependent incoming resource ratios in scale-free networks. Nodes are grouped into four degree-quantile groups, $Q_1$--$Q_4$, according to the empirical degree distribution. For the $N=50$ BA network analyzed here, the groups correspond to $d=2$, $d=3$,$d=4$--$6$, and $d=7$--$19$, respectively.
    Each trajectory is divided into three temporal stages: the first 20\%, the middle 60\%, and the last 20\% of the episode.
    For each stage, nodes are grouped by receiver degree into four bins.
    The red curve shows the ratio of observed incoming resources under the learned RL agent to those under the equal baseline.
    The blue curve shows the ratio of observed incoming resources under the learned RL agent to those under the proportional baseline.
    A ratio larger than 1 indicates that the corresponding degree bin receives more resources than the baseline, whereas a ratio smaller than 1 indicates that it receives less. (d)-(g): degree-conditioned mixture mechanism M2 in scale-free networks. The four panels correspond to different focal-node degree bins: Q1:2, Q2:3, Q3:4--6, and Q4:7--19.
    For each degree bin, the horizontal axis represents the ego-network average accumulated resource $\bar{R}_{\mathrm{ego}}(t)$, and the vertical axis represents the mixture weight of each allocation component.
    The three components are proportional allocation, equal allocation, and self-allocation.
    Circular markers denote binned empirical weights estimated from the RL-agent rollout, while solid step curves denote the fitted discrete mechanism used by M2.}
    \label{fig:scale_free_diagram}
\end{figure}
We next examine how the learned allocation mechanism behaves when structural heterogeneity is introduced. For this purpose, we focus on scale-free networks, where node degrees are highly heterogeneous, and a small number of hub nodes coexist with many low-degree nodes. This setting allows us to investigate whether the learned planner treats individuals differently according to their structural positions.

To quantify the degree-dependent allocation pattern, we divide each trajectory into three temporal stages: the first 20\% (steps 1--40), the middle 60\% (steps 41--160), and the last 20\% (steps 161--200) of the episode. Within each stage, we compute the accumulated incoming resources received by each node. Nodes are then grouped according to their receiver degree into four bins: Q1:2, Q2:3, Q3:4--6, and Q4:7--19. For each bin, we report how much resource the learned RL-agent allocates relative to the Equal and Proportional baselines.
A ratio larger than 1 indicates that the corresponding degree bin receives more resources than the baseline prediction, whereas a ratio smaller than 1 indicates that it receives less. The analysis is based on 100,000 node-level observations collected from the frozen RL policy on a 50-node BA network.
As shown in Fig.~\ref{fig:scale_free_diagram}(a)-(c), the observed allocation differs strongly from the equal baseline. Low-degree nodes receive substantially more resources than they would under equal allocation, while high-degree nodes receive much less. This suggests that the learned planner does not simply amplify the structural advantage of high-degree nodes, even though such nodes participate in more local pools.
By contrast, the observed allocation is much closer to the proportional baseline. The ratio relative to proportional allocation remains close to 1 across degree bins and temporal stages, with only mild deviations for higher-degree nodes. This indicates that, in heterogeneous degree networks, the learned policy preserves a contribution-related allocation structure while adjusting it according to degree heterogeneity.
Importantly, this degree-dependent pattern is relatively stable across the first, middle, and last stages of the episode. Therefore, the degree effect should be interpreted mainly as a structural feature of the learned allocation mechanism, rather than as a transient temporal effect. This motivates the following degree-conditioned analysis, in which
structural heterogeneity is explicitly incorporated into the
interpretable mixture mechanism. Since allocation decisions are made at the resource-pool level, M2 conditions the mixture weights on the degree of the focal node defining each local pool.

Based on the above degree-dependent analysis, we further construct a degree-conditioned mixture mechanism for heterogeneous networks, denoted by M2. Compared with M1, where the mixture weights depend only on the ego-network average accumulated resource, M2 additionally conditions the mixture weights on the degree of the focal node. This extension is motivated by the observation that, in scale-free networks, nodes with different degrees play different structural roles and may require different local allocation mechanisms.

For each resource pool centered at node $i$, we use the ego-network average accumulated resource $\bar{R}_{\mathrm{ego}, i}(t)$ to characterize the local resource state, and use the degree $d_i$ of the focal node to characterize its structural role. The focal nodes are grouped into four degree-quantile groups, $Q_1$--$Q_4$, according to the empirical degree distribution of the network. Thus, the grouping is defined by relative structural position rather than by fixed degree thresholds. In the $N=50$ BA network used for policy distillation, these four quantile groups correspond to
$Q_1:d_i=2$, $Q_2:d_i=3$, $Q_3:d_i=4$--$6$, and
$Q_4:d_i=7$--$19$, respectively. Within each degree bin, we fit a resource-dependent mixture mechanism over three representative allocation components: proportional allocation, equal allocation, and self-allocation.
Formally, the M2 allocation proportion from pool $i$ to individual $j\in N_i\cup\{i\}$ is written as
\[
\hat{A}_{i\to j}^{\mathrm{M2}}(t)
=
\omega_{\mathrm{Prop}}^{(q_i,b_i(t))}
\frac{e_j(t)/(d_j+1)}
{\sum_{k\in N_i\cup\{i\}} e_k(t)/(d_k+1)}
+
\omega_{\mathrm{Equal}}^{(q_i,b_i(t))}
\frac{1}{d_i+1}
+
\omega_{\mathrm{Self}}^{(q_i,b_i(t))}
\mathbf{1}_{j=i},
\]
where $q_i$ denotes the degree bin of the focal node $i$, and $b_i(t)$ denotes the resource bin determined by $\bar{R}_{\mathrm{ego},i}(t)$. For each pair of bins $(q_i,b_i(t))$, the three mixture weights are non-negative and sum to one.
Fig.~\ref{fig:scale_free_diagram}(d)-(g) visualizes the fitted degree-conditioned mixture mechanism. The four panels correspond to different focal-node degree bins. Within each panel, the horizontal axis represents the ego-network average accumulated resource $\bar{R}_{\mathrm{ego}}(t)$, and the vertical axis represents the fitted mixture weight of each allocation component. The circular markers show the binned empirical weights estimated from the RL-agent rollout, while the solid step curves show the discrete mechanism used by M2.

The fitted profiles reveal that the learned allocation mechanism changes substantially across degree groups. For low-degree focal nodes, the self-allocation component can become important in some intermediate resource regimes. This indicates that the learned planner tends to preserve resources around structurally peripheral nodes. In contrast, for middle-degree nodes, the proportional component dominates almost all resource regimes, suggesting that contribution-based allocation is sufficient for these nodes. However, when total resources are low, a certain equal baseline coefficient is applied to prevent individuals from falling into the poverty trap. For high-degree focal nodes, the equal component becomes more pronounced when the ego-network resource level is high, indicating that the planner introduces stronger redistribution around hub-like local pools. These results suggest that, in heterogeneous networks, the learned policy cannot be fully captured by a resource-only mechanism. Instead, it is better interpreted as a degree-conditioned mixture mechanism that adapts allocation principles according to both local resource states and structural roles. Subsequent simulations further demonstrated this point.
\begin{figure}[!htbp]
    \centering
    \includegraphics[width=0.95\linewidth]{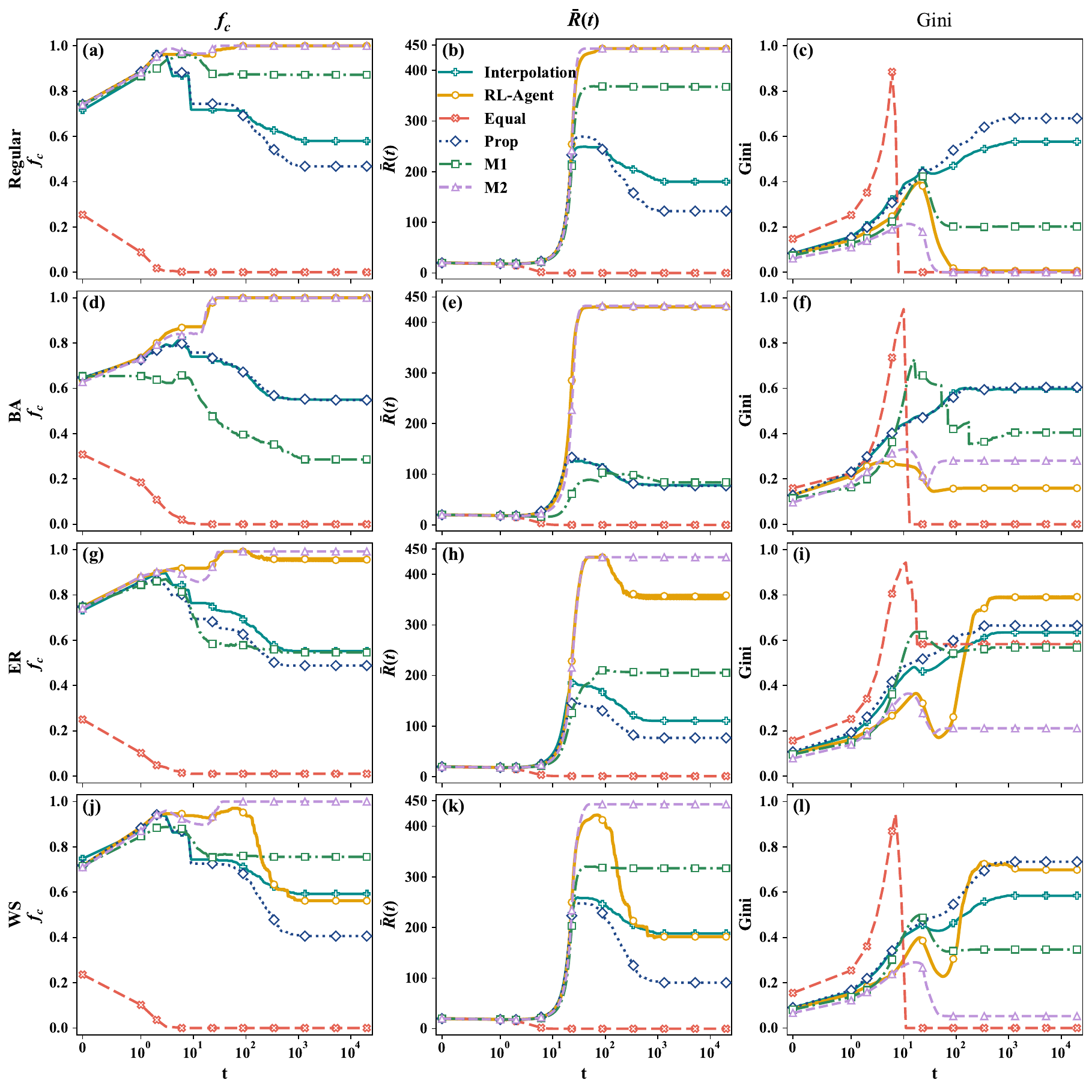}
    \caption{Evaluation metrics over time in four topologies under different allocation mechanisms. Rows correspond to  (a)-(c): Regular, (d)-(f): BA, (g)-(i): ER, and (j)-(l): WS networks. Columns correspond to \(f_c\), \(\bar{R}(t)\), and $Gini$, respectively. The compared mechanisms are M1, M2, Proportional (Prop), Equal, Interpolation, and the RL-Agent. The horizontal axis shows time \(t\) on a logarithmic scale ($T=2\times10^4$). The results are obtained by averaging 50 independent simulations.}
    \label{fig:five_policy_diagram_7}
\end{figure}

% 分析图中表现，BA下M2的Gini倒是不如RL-Agent，会稍差一点点。

To further evaluate the performance of the interpretable mixture mechanisms M1 and M2, we tested six allocation mechanisms on newly generated random graph instances in Fig.~\ref{fig:five_policy_diagram_7}. Unlike the evaluation conducted during training, the graph instances used here are fully resampled and are not taken from the fixed topology bank used during RL training. This setting provides a stricter test of generalization to unseen network structures. In addition to the randomly sampled network evaluation reported here, we further test the six mechanisms under parameter shifts in $p_c$, $\kappa$, $r$ and $N$ (see~\ref{Appendix:Additional Generalization Tests}).
Fig.~\ref{fig:five_policy_diagram_7} compares the equal baseline, the proportional baseline, the Interpolation baseline, the RL-Agent, the resource-binned mixture mechanism M1, and the degree-conditioned mixture mechanism M2 across four network topologies. The three columns report the cooperation level $f_c(t)$, the average accumulated resource $\bar{R}(t)$, and the Gini coefficient, respectively.
The Equal baseline rapidly collapses in all network topologies, confirming that purely uniform redistribution cannot maintain cooperation. The Proportional baseline performs better than the Equal baseline at the early stage, but it eventually leads to declining cooperation, reduced accumulated resources, and high inequality, especially in heterogeneous networks.
The resource-binned mechanism M1 performs well on regular networks, where degree heterogeneity is absent. However, its performance deteriorates on BA, ER, and WS networks. This result is consistent with the previous interpretation: M1 only conditions the allocation mechanism on the ego-network resource level and therefore cannot fully account for structural heterogeneity.
The original RL-Agent also exhibits strong performance on several topologies, but its dynamics become less stable on unseen random graphs. In particular, under ER and WS networks, the RL-Agent initially improves cooperation and accumulated resources, but the performance later declines, accompanied by a marked increase in inequality.
This suggests that the RL-Agent is sensitive to distributional changes in graph structure and to the long-term stochastic evolutionary process when evaluated on more diverse network instances and over longer evolutionary horizons than those used during training.

By contrast, M2 achieves the most robust overall performance across all four topologies. It maintains a high cooperation level and high accumulated resources in regular, BA, ER, and WS networks, while keeping the Gini coefficient relatively low. This indicates that explicitly incorporating the degree-conditioned allocation mechanism improves the generalization of the learned mechanism. In other words, M2 captures not only the resource-dependent pattern identified in regular networks, but also the structural adjustment required in degree-heterogeneous networks.

To further interpret the results, we compare the observed Gini coefficients with the theoretical and simulation benchmarks under the full-cooperation equal-allocation conditions derived in~\ref{Appendix:Analysis of Steady State}. These benchmarks describe the level of resource inequality induced purely by degree heterogeneity under full cooperation and equal local redistribution. Under the default parameter setting, the corresponding Gini theoretical benchmarks are approximately $0.3117$ for BA networks, $0.2122$ for ER networks, and $0.0564$ for WS networks. Furthermore, we conducted 50 simulations under the same conditions to obtain specific Gini index data at the steady state of persistent cooperation: BA: 0.2842, ER: 0.2111, and WS: 0.0529.
Interestingly, M2 achieves Gini coefficients of approximately $0.2805$, $0.2111$, and $0.0525$ on BA, ER, and WS networks, respectively, while maintaining high cooperation levels and near-optimal average accumulated resources. This suggests that the M2 strategy not only promotes cooperation and increases accumulated resources, but also maintains a level of fairness that is almost equivalent to that of equal distribution.
% In particular, the improvement is most evident in the BA scale-free network, where degree heterogeneity is strongest. This result suggests that equal allocation does not necessarily imply global fairness in heterogeneous networks. Even if every local pool distributes resources equally, nodes with different degrees participate in different numbers of pools, which can generate systematic inequality at the population level. This also indicates that purely equal allocation is not necessarily globally fair in degree-heterogeneous networks. Instead, a structure-aware allocation mechanism can compensate for degree-induced inequality by adapting redistribution to network position.
The degree-conditioned structure of M2 provides an interpretable explanation for this result. For low-degree focal nodes, M2 assigns a higher weight to the self-allocation component in some resource regimes, which helps peripheral nodes preserve their own resources and maintain the ability to continue cooperating. For middle-degree nodes, the proportional component dominates, indicating that contribution-based incentives are sufficient to sustain cooperative development. At the same time, M2 maintain a certain degree of equal baseline allocation when ego-network resources are scarce to prevent individuals from falling into the poverty trap. For high-degree nodes, the equal-allocation component becomes more pronounced, suggesting that hub-like nodes should place more emphasis on supporting their surrounding neighbours rather than only reinforcing their own contribution advantage. Therefore, M2 can be interpreted as a structure-aware allocation mechanism: peripheral nodes are protected from falling into poverty traps, middle-degree nodes are incentivized through proportional rewards, and high-degree nodes help redistribute resources within their larger local neighbourhoods.

\section{Discussion and Implications}

\subsection{Theoretical implications}

This study provides several theoretical implications for cooperation research, networked resource allocation, and learning-based mechanism design. First, the results show that allocation mechanisms are not merely outcomes of cooperation, but can actively shape the future capacity of individuals to cooperate. In the proposed network common-pool resource game, the resources received by an individual affect its subsequent ability to contribute. Therefore, allocation rules influence not only immediate payoffs, but also the long-term feedback loop among resource accumulation, contribution capacity, and strategy evolution.
Second, the findings reveal a tension between local fairness and global fairness in networked resource-sharing systems. Equal allocation appears fair within each local pool, but it may remove contribution incentives and drive the system toward cooperation collapse. In heterogeneous networks, even persistent cooperation and equal local redistribution may still generate nonzero resource inequality, because individuals with different degrees participate in different numbers of local pools. This suggests that fairness in networked common-pool systems cannot be evaluated only at the level of a single interaction group; it must also account for structural positions in the whole network.
Third, this study shows that deep reinforcement learning can be used not only as an optimization tool, but also as a mechanism-discovery tool. 
Accordingly, the emphasis here is on the allocation mechanisms discovered from the learned policy rather than on the superiority of a particular reinforcement learning algorithm.
The GNN-RL social planner learns an adaptive allocation policy that balances cooperation, resource sustainability, and inequality. More importantly, by interpreting and distilling the learned policy, the black-box neural policy can be transformed into simpler resource-dependent and degree-conditioned mixture mechanisms. This provides an interpretable route from learning-based policy search to mechanism-design principles in networked social systems.

\subsection{Practical implications}

The results also suggest potential practical implications for public resource allocation, platform governance, ecological management, and organizational incentive design. In many real systems, resources are repeatedly allocated to individuals or groups whose subsequent behaviour determines whether the system can remain sustainable. The findings suggest that purely equal allocation may be insufficient when it weakens incentives to contribute, while purely contribution-proportional allocation may become problematic when it amplifies early advantages and pushes resource-poor individuals into persistent disadvantage.
Within the modeled setting, adaptive allocation mechanisms perform more favorably than fixed allocation rules, suggesting that state-dependent allocation may be worth considering in related real-world systems.
The simulations suggest that, when local resources are scarce, some protective redistribution can help prevent persistent resource disadvantage. When individuals are in intermediate structural positions, contribution-based rewards can help maintain incentives. In the modeled heterogeneous networks, greater redistribution around highly connected pools is associated with lower resource concentration. These principles suggest that sustainable governance in networked systems may benefit from allocation rules that are sensitive to both local scarcity and network position.
For AI-assisted decision systems, the study further suggests that artificial intelligence should not only be evaluated by aggregate performance, but also by whether its intervention principles can be understood and inspected. The extracted mixture mechanisms provide a more transparent alternative to directly deploying a black-box neural planner. This is particularly important when AI systems are used to support decisions related to social welfare, resource allocation, or collective governance, where interpretability and alignment with human welfare are essential.

\subsection{Limitations and future work}

This study has several limitations. First, the results are based on simulation-generated data rather than empirical human-subject or real-world behavioural data. Although the proposed model captures important features of networked common-pool resource systems, such as overlapping local pools, endogenous resource constraints, and heterogeneous structural positions, real-world resource-sharing systems may involve richer institutional rules, behavioural heterogeneity, and contextual factors. Future work could evaluate the proposed mechanisms using empirical networks or controlled behavioural experiments.
Second, a fixed Fermi rule governs individual strategy updates in the current model. This assumption allows us to focus on the effect of allocation mechanisms, but it simplifies human decision-making behaviour. In reality, individuals may use bounded rationality, learning, social norms, reciprocity, or heterogeneous preferences when deciding whether to cooperate. Future research could introduce human participants, human-behaviour models, or heterogeneous learning agents to test whether the extracted allocation mechanisms remain effective under more realistic behavioural assumptions \cite{jia2025social,mckee2023scaffolding,koster2025deep}.
Third, the current framework considers static network structures during each simulation. Many real social, economic, and ecological systems involve dynamic networks in which individuals can form, maintain, or sever connections over time. Extending the framework to coevolving networks would make it possible to study how allocation mechanisms and network adaptation jointly influence cooperation. In addition, future work could examine larger networks, changing resource environments, and multi-level governance structures. Future work could also formulate cooperation, efficiency, and fairness as
explicit multiple objectives or policy constraints, allowing the corresponding trade-offs to be explored directly through multi-objective or constrained reinforcement learning.
Finally, although the learned policy is distilled into interpretable mixture mechanisms, the theoretical stability and optimality of these extracted mechanisms remain to be further analysed. Future research may derive more explicit analytical conditions under which resource-dependent and degree-conditioned allocation mechanisms can sustain cooperation, reduce inequality, and prevent resource collapse. More broadly, when AI systems are deployed as social planners in real-world settings, their interventions should be evaluated not only by efficiency, but also by transparency, fairness, robustness, and alignment with human welfare \cite{summerfield2024will,hackenburg2026artificial,schroeder2026malicious}.

\section{Conclusions}

This study investigated how adaptive allocation mechanisms can promote cooperation in resource-constrained networked populations. A network common-pool resource game was proposed, in which individuals are embedded in complex networks, participate in overlapping local resource pools, and face endogenous resource constraints during strategy evolution. This framework captures the coupling among cooperation, resource accumulation, local interaction structure, and allocation rules.
The results show that simple hand-designed allocation mechanisms face inherent limitations. Equal allocation may produce superficially fair outcomes but weakens contribution incentives and can lead to cooperation collapse. Proportional allocation can temporarily stimulate cooperation, but it may push resource-poor individuals into poverty traps and amplify accumulated advantages. 
By contrast, the proposed GNN-RL social planner learns adaptive allocation policies that achieve more balanced outcomes across different network topologies by sustaining higher cooperation levels, maintaining larger average accumulated resources, and reducing inequality.
More importantly, the learned policy reveals interpretable principles for resource sharing. On regular networks, the policy can be approximated by a resource-dependent mixture mechanism. On heterogeneous networks, the degree-conditioned mixture mechanism further shows that peripheral nodes, intermediate-degree nodes, and hub-like nodes require different combinations of protection, contribution-based reward, and redistribution. 
Within the modeled network common-pool resource setting, these findings indicate that successful allocation mechanisms can benefit from conditioning redistribution on both local resource states and structural positions.
In conclusion, this study integrates explainable AI methods and network science to uncover the cooperative mechanisms of networked social systems.

\section*{Declaration of competing interest}
The authors declare that they have no known competing financial interests or personal relationships that could have appeared to influence the work reported in this paper.
\section*{Acknowledgments}
This work was supported in part by the National Natural Science Foundation of China under Grants 62373245 and 12426311,
and by the National Key R\&D Program of China under Grant 2023YFB4706800, 
and in part by the “Dawn” Program of Shanghai Education Commission, China.
\section*{Data availability}
No data was used for the research described in the article
\appendix
\setcounter{figure}{0}
\renewcommand{\thefigure}{A.\arabic{figure}}
\setcounter{table}{0}
\renewcommand{\thetable}{A.\arabic{table}}

\section{Implementation details of the GNN-RL planner}
\label{Appendix:Network Architecture Specification}
\setcounter{equation}{0}
\renewcommand{\theequation}{A.\arabic{equation}}

This appendix provides additional implementation details of the proposed GNN-RL planner. 
The planner is designed to act as an external social planner that observes the current networked resource state and determines how the allocable resources of each local pool should be distributed. 
Its implementation consists of two main parts: a graph-based actor network that outputs local allocation weights and twin graph-based critic networks that estimate the action value of the allocation policy.

The detailed network architecture is summarized in Table~\ref{tab:architecture_summary}. 
The actor first encodes the global network state through stacked GraphNet blocks and then applies a shared ego-network allocation head to each focal node. 
This design allows the same local allocation rule to be repeatedly applied to different resource pools while still conditioning the decision on the local neighborhood structure and node-level resource states. 
The critic networks use a similar graph-based state representation but are independently parameterized from the actor, so that policy evaluation and policy improvement remain separated during training.

\begin{table}[htbp]
\centering
\begin{threeparttable}
\caption{Summary of the network architecture used in the proposed GNN-RL planner.}
\label{tab:architecture_summary}
\begin{tabular}{p{3.8cm} p{8.8cm}}
\toprule
Component & Specification \\
\midrule
Input representation 
& Graph tensor \((u,V,E)\), where \(u\) denotes global features, \(V\) denotes node features, and \(E\) denotes edge features. \\

First GraphNet block 
& MLP-based message passing updates for edge, node, and global representations; hidden dimension 128. \\

Second GraphNet block 
& MLP-based message passing updates for edge, node, and global representations; hidden dimension 128. \\

Ego-network extraction 
& For each focal node, an ego-network consisting of the node and its one-hop neighbours is extracted as a local resource pool. \\

Local ego-network head 
& A shared local GraphNet block is applied to each ego-network; hidden dimension 64. \\

Allocation score head 
& An MLP-based score readout with hidden dimension 64, followed by softmax normalization to produce row-wise allocation weights. \\

Actor output 
& Row-wise local allocations are assembled into the final allocation action over all local resource pools. \\

Twin critics 
& Two graph action-value critics with identical architecture and independent parameters. \\

Critic state encoder 
& Two stacked GraphNet blocks with hidden dimension 64. \\

Critic value head 
& MLP-based local action encoder, pool-level encoder, and scalar \(Q\)-value head, all with hidden dimension 64. \\

Activation and sharing 
& GELU is used in MLP modules; the ego-network allocation head is shared across all focal nodes. \\
\bottomrule
\end{tabular}
\begin{tablenotes}
\footnotesize
\item Notes: (1) The actor global encoder and the critic state encoder use the same input format but do not share parameters.
(2) The ego-network allocation head is applied repeatedly to every focal node with shared weights.
(3) The twin critics share the same architecture but are independently parameterized.
\end{tablenotes}
\end{threeparttable}
\end{table}

The training hyperparameters are reported in Table~\ref{tab:main_training_parameters}. 
We use the TD3-style training scheme with twin critics, delayed actor updates, and soft target-network updates. 
These components are used to reduce overestimation bias and improve training stability in the non-stationary evolutionary environment. 
During training, each episode contains 200 evolutionary steps, while the evaluation experiments in the main text use longer simulation horizons to examine whether the learned allocation mechanism can sustain cooperation over extended dynamics. 
The social planner's reward at each step is defined as the next-step average accumulated resource.
At each training step, after the planner determines the allocation action and the environment evolves to the next state, the planner receives a scaled next-step average-resource reward:
\begin{equation}
r_{\mathrm{planner}}(t)
=
\lambda_R \bar{R}(t+1).
\end{equation}
Here, \(\lambda_R>0\) is a constant reward-scaling coefficient used to adjust the numerical magnitude of the reward during training.
It should be noted that the reward does not explicitly include the cooperation level or the Gini coefficient. 
Therefore, the cooperation and fairness outcomes reported in the main text are not directly optimized objectives. In particular, a low Gini coefficient is not mathematically guaranteed by the resource-oriented reward, as high average accumulated resources can coexist with substantial inequality. The observed fairness improvements of the learned and distilled mechanisms should therefore be interpreted as empirical properties of the resulting allocation policies. Meanwhile, sustained resource accumulation depends on continued cooperative replenishment through the network dynamics.

\begin{table}[htbp]
\centering
\caption{Main training hyperparameters of the proposed GNN-RL planner.}
\label{tab:main_training_parameters}
\begin{tabular}{lll}
\toprule
Parameter & Description & Value \\
\midrule
Training steps & Total number of environment steps & \(5\times10^7\) \\
Episode length ($T_E$) & Number of steps per training episode & \(200\) \\
Batch size & Mini-batch size for replay-buffer sampling & \(256\) \\
Actor learning rate & Learning rate of the actor network & \(5\times10^{-6}\) \\
Critic learning rate & Learning rate of the critic networks & \(2\times10^{-5}\) \\
\(\gamma\) & Discount factor & \(0.99\) \\
\(\rho\) & Target-network soft update rate & \(0.005\) \\
Policy delay & Frequency of delayed actor updates & \(8\) \\
Gradient steps & Gradient steps per environment update & \(2\) \\
Reward signal & Scaled next-step average-resource reward & \(\lambda_R \bar{R}(t+1)\) \\
\(\lambda_R\) & Reward scaling coefficient & 0.027 \\
\bottomrule
\end{tabular}
\end{table}

\section{Implementation details of mixture allocations}
\subsection{resource-binned mixture allocation mechanism}
% \section{Fitting and reconstruction accuracy of M1}
\label{Appendix:M1_fitting}
\setcounter{equation}{0}
\renewcommand{\theequation}{B.\arabic{equation}}
The resource-dependent mixture mechanism M1 is obtained through a post-hoc behavioural decomposition of the frozen RL policy. For each ego-network state associated with the resource pool centered at node $i$, the RL actor produces an allocation vector
\begin{equation}
\mathbf{A}^{\mathrm{RL}}_{i}(t)
=
\left(
\hat{A}^{\mathrm{RL}}_{i\to j}(t)
\right)_{j\in N_i\cup\{i\}},
\end{equation}
where the allocation proportions are non-negative and sum to one.

For the same ego-network state, we construct three reference allocation vectors corresponding to proportional allocation, equal allocation, and self-allocation, denoted by
$\mathbf{A}^{\mathrm{Prop}}_{i}(t)$,
$\mathbf{A}^{\mathrm{Equal}}_{i}(t)$, and
$\mathbf{A}^{\mathrm{Self}}_{i}(t)$,
respectively. The RL allocation is approximated as
\begin{equation}
\widehat{\mathbf{A}}_{i}(t)
=
w_{\mathrm{Prop}}\mathbf{A}^{\mathrm{Prop}}_{i}(t)
+
w_{\mathrm{Equal}}\mathbf{A}^{\mathrm{Equal}}_{i}(t)
+
w_{\mathrm{Self}}\mathbf{A}^{\mathrm{Self}}_{i}(t),
\end{equation}
where the coefficients are estimated by least-squares fitting to minimize the squared $L_2$ reconstruction error. The fitted coefficient vector is subsequently projected onto the probability simplex such that
\begin{equation}
w_{\mathrm{Prop}},w_{\mathrm{Equal}},w_{\mathrm{Self}}\geq 0,
\qquad
w_{\mathrm{Prop}}+w_{\mathrm{Equal}}+w_{\mathrm{Self}}=1.
\end{equation}

The fitting procedure was applied to 40,000 ego-network states collected from episodes of the frozen RL policy on regular networks, where each episode contains 200 evolutionary steps. The mean absolute reconstruction error of the state-level three-component decomposition was $1.95\times10^{-3}$ per individual in the ego-network, indicating that the three reference mechanisms provide a close behavioural approximation to the allocation outputs of the RL actor.

To construct the resource-dependent M1 mechanism, the fitted state-level weights are grouped according to the mean accumulated resource of the corresponding ego-network,
\begin{equation}
\bar{R}_{\mathrm{ego},i}(t)
=
\frac{1}{d_i+1}
\sum_{j\in N_i\cup\{i\}}R_j(t).
\end{equation}
The ego-network mean resource was discretized using quantile-based binning. Starting from 20 equal-frequency quantile intervals, repeated upper boundaries caused by the resource cap at 443.18 were merged, resulting in 14 effective resource bins.
The M1 coefficients associated with each bin are defined as the arithmetic mean of the fitted state-level coefficients within that bin. The resulting bin boundaries and mixture weights are reported in Table~\ref{tab:M1_weights}.

We further evaluate how accurately the final binned M1 mechanism reproduces the allocation behaviour of the RL policy. Importantly, this evaluation uses the fixed bin-level weights shown in Fig.~\ref{fig:regular_diagram}(c), rather than re-fitting a separate set of coefficients for each ego-network state. For a state belonging to resource bin $b$, the reconstructed M1 allocation is
\begin{equation}
\mathbf{A}^{\mathrm{M1}}_{i}(t)
=
\bar{\omega}_{\mathrm{Prop}}^{(b)}
\mathbf{A}^{\mathrm{Prop}}_{i}(t)
+
\bar{\omega}_{\mathrm{Equal}}^{(b)}
\mathbf{A}^{\mathrm{Equal}}_{i}(t)
+
\bar{\omega}_{\mathrm{Self}}^{(b)}
\mathbf{A}^{\mathrm{Self}}_{i}(t).
\end{equation}

The reconstruction error for an ego-network state is measured by the mean absolute difference between the RL and M1 allocation proportions,
\begin{equation}
E_{i,t}^{\mathrm{M1}}
=
\frac{1}{d_i+1}
\sum_{j\in N_i\cup\{i\}}
\left|
\hat{A}^{\mathrm{RL}}_{i\to j}(t)
-
\hat{A}^{\mathrm{M1}}_{i\to j}(t)
\right|.
\end{equation}

Across all 40,000 ego-network states, the final binned M1 mechanism achieves an overall mean absolute reconstruction error of
\begin{equation}
\overline{E}^{\mathrm{M1}}=0.00678.
\end{equation}
The corresponding allocation-entry root mean squared error is $0.02568$. Although the use of a single set of weights within each resource bin increases the error relative to the state-specific decomposition, the absolute reconstruction error remains small overall. The largest deviation occurs in the lowest-resource bin, $\bar{R}_{\mathrm{ego}}\in[0,17.043)$, where the mean absolute error is $0.07011$. In contrast, the mean absolute reconstruction error is below $0.01$ in most of the remaining resource bins. These results show that the resource-binned M1 mechanism retains a close approximation to the allocation behaviour of the original RL policy while substantially reducing the policy to a small set of interpretable, resource-dependent mixture weights.
All bin-level design matrices remained full rank. The condition numbers ranged from 1.53 to 24.16, with larger values occurring mainly in the high-resource regime, where proportional allocation increasingly approaches equal allocation.
\begin{table}[htbp]
\centering
\caption{Fitted M1 profiles.}
\label{tab:M1_weights}
\small
\setlength{\tabcolsep}{4pt}
\begin{tabular}{ccccccc}
\hline
Resource bin &
Mean $\bar{R}_{\mathrm{ego}}$ &
$\omega_{\mathrm{Prop}}$ &
$\omega_{\mathrm{Equal}}$ &
$\omega_{\mathrm{Self}}$ &
MAE &
$\kappa_2(X)$ \\
\hline
$[0,17.043)$
& 9.010
& 0.5539 & 0.4046 & 0.0415
& 0.07011 & 1.5789 \\

$[17.043,24.258)$
& 19.844
& 0.9482 & 0.0449 & 0.0069
& 0.00817 & 1.5270 \\

$[24.258,53.379)$
& 37.307
& 0.9871 & 0.0099 & 0.0030
& 0.00313 & 1.7257 \\

$[53.379,103.593)$
& 74.167
& 0.9794 & 0.0179 & 0.0028
& 0.00260 & 1.6088 \\

$[103.593,184.283)$
& 141.903
& 0.9691 & 0.0257 & 0.0052
& 0.00424 & 1.7620 \\

$[184.283,382.775)$
& 292.098
& 0.9592 & 0.0244 & 0.0164
& 0.00930 & 1.7500 \\

$[382.775,431.676)$
& 415.775
& 0.9064 & 0.0738 & 0.0198
& 0.00640 & 3.8641 \\

$[431.676,439.884)$
& 437.770
& 0.8780 & 0.1086 & 0.0134
& 0.00300 & 6.2303 \\

$[439.884,440.581)$
& 440.106
& 0.8679 & 0.1208 & 0.0113
& 0.00063 & 18.6338 \\

$[440.581,441.384)$
& 440.800
& 0.8653 & 0.1232 & 0.0115
& 0.00074 & 16.9713 \\

$[441.384,441.845)$
& 441.653
& 0.8878 & 0.1033 & 0.0089
& 0.00140 & 20.4307 \\

$[441.845,442.487)$
& 442.144
& 0.9683 & 0.0287 & 0.0030
& 0.00289 & 16.0235 \\

$[442.487,442.913)$
& 442.693
& 0.9233 & 0.0705 & 0.0061
& 0.00186 & 24.1595 \\

$[442.913,443.180]$
& 451.993$^{\dagger}$
& 0.9308 & 0.0627 & 0.0065
& 0.00302 & 8.6015 \\
\hline
\end{tabular}

\vspace{0.5em}
\begin{minipage}{0.96\linewidth}
\footnotesize
MAE denotes the mean absolute reconstruction error between the RL allocation
and the final binned M1 allocation. $\kappa_2(X)$ denotes the spectral
condition number of the bin-level least-squares design matrix
$X=[\mathbf{A}^{\mathrm{Prop}}-\mathbf{A}^{\mathrm{Equal}},
\mathbf{A}^{\mathrm{Self}}-\mathbf{A}^{\mathrm{Equal}}]$.
$^{\dagger}$The original mean ego-network resource in the final bin is 451.993.
For bin assignment, resource values above 443.180 are capped at 443.180,
giving an effective mean of 443.178 for this bin.
\end{minipage}
\end{table}

\subsection{Degree-conditioned mixture allocation mechanism}
\label{Appendix:M2_fitting}

The degree-conditioned mixture mechanism M2 extends M1 by incorporating network structural heterogeneity into the distilled allocation rule. While M1 conditions the mixture weights only on the ego-network mean accumulated
resource, M2 jointly conditions them on the degree of the focal node and the resource state of the corresponding ego-network.
The state-level behavioural decomposition follows the same procedure as described in~\ref{Appendix:M1_fitting}. Specifically, for each ego-network state, the allocation output of the frozen RL actor is decomposed into proportional allocation, equal allocation, and self-allocation using
least-squares fitting followed by projection onto the probability simplex. Therefore, the fitting objective and the definitions of the three reference allocation mechanisms are not repeated here.
The M2 fitting procedure was applied to 50,000 ego-network states collected from episodes of the frozen RL policy on a Barab\'asi--Albert(BA) scale-free network with $N=50$, where each episode contains 200 evolutionary steps. 
To characterize the structural role of each local
resource pool, the focal nodes were first ranked according to degree and grouped into four degree-quantile groups, denoted by $Q_1$--$Q_4$, according to the empirical degree distribution. The degree quantiles therefore represent relative structural positions in the network rather than fixed absolute
degree intervals. Because degree is discrete, identical degree values were kept within the same group. For the $N=50$ BA network used in the present policy-distillation analysis, the resulting groups correspond to
$Q_1:d_i=2$, $Q_2:d_i=3$, $Q_3:d_i=4$--$6$, and
$Q_4:d_i=7$--$19$.

Within each degree group, the ego-network mean accumulated resource $\bar{R}_{\mathrm{ego},i}(t)$ was further discretized using quantile-based binning. Twelve quantile intervals were initially specified for each degree
group. Resource values above 430 were assigned to the terminal resource level, and repeated quantile boundaries were merged when necessary.
Consequently, $Q_1$ contains 10 effective resource bins, whereas $Q_2$--$Q_4$ each contain 12 effective resource bins.
For each degree--resource bin, the final M2 coefficients are defined as the arithmetic mean of the state-level fitted mixture weights within that bin.
Thus, each combination of degree group and resource bin is associated with a fixed triplet of proportional, equal, and self-allocation weights. The solid step curves in Fig.~\ref{fig:scale_free_diagram}(d)--(g) represent
these final discrete M2 weights.
The mean absolute reconstruction error of the state-level three-component decomposition is $7.46\times10^{-3}$ per individual in the ego-network.
We further evaluated the final degree-conditioned M2 mechanism using the fixed degree--resource-bin weights, without re-fitting state-specific coefficients. Across all 50,000 ego-network states, the final M2 mechanism achieves an overall mean absolute reconstruction error of
$0.01261$.
The reconstruction error is slightly larger for the Q1 group, whose overall MAE is $0.01765$, whereas the corresponding values for
$Q_2$, $Q_3$, and $Q_4$ are $0.00885$, $0.00715$, and $0.00752$, respectively. The largest bin-level deviation occurs in the Q1 group at intermediate resource levels, while the remaining degree groups exhibit relatively small reconstruction errors across most resource regimes.

We additionally examined the numerical conditioning of the decomposition using the same reduced two-direction design matrix as in~\ref{Appendix:M1_fitting}. The bin-level spectral condition numbers range from $1.58$ to $11.06$, indicating that none of the degree--resource
bins are severely ill-conditioned at the aggregated bin level. Detailed results are reported in Tables~\ref{tab:M2_Q1}--\ref{tab:M2_Q4}.

\begin{table}[htbp]
\centering
\caption{Fitted M2 profiles for degree group $Q_1:d_i=2$.}
\label{tab:M2_Q1}
\small
\setlength{\tabcolsep}{3.5pt}
\begin{tabular}{ccccccc}
\hline
Resource bin &
Mean $\bar{R}_{\mathrm{ego}}$ &
$\omega_{\mathrm{Prop}}$ &
$\omega_{\mathrm{Equal}}$ &
$\omega_{\mathrm{Self}}$ &
MAE &
$\kappa_2(X)$ \\
\hline
$[2.636,67.111)$     & 27.802  & 0.7599 & 0.0990 & 0.1411 & 0.03879 & 3.1692 \\
$[67.111,252.776)$   & 162.910 & 0.5137 & 0.0173 & 0.4691 & 0.05394 & 5.4391 \\
$[252.776,293.296)$  & 278.776 & 0.8260 & 0.0040 & 0.1700 & 0.00549 & 11.0612 \\
$[293.296,319.700)$  & 305.500 & 0.9116 & 0.0019 & 0.0865 & 0.01745 & 4.5876 \\
$[319.700,322.386)$  & 320.930 & 0.9572 & 0.0005 & 0.0423 & 0.00882 & 5.2958 \\
$[322.386,334.624)$  & 326.645 & 0.8986 & 0.0003 & 0.1010 & 0.00525 & 9.2479 \\
$[334.624,346.817)$  & 341.665 & 0.9891 & 0.0001 & 0.0108 & 0.00526 & 3.3489 \\
$[346.817,371.641)$  & 358.815 & 0.9948 & 0.0002 & 0.0050 & 0.00606 & 4.1268 \\
$[371.641,392.124)$  & 373.932 & 0.9897 & 0.0001 & 0.0102 & 0.00837 & 4.9149 \\
$[392.124,430.000]$  & 444.230$^{\dagger}$ & 0.8703 & 0.1182 & 0.0116 & 0.02078 & 4.6518 \\
\hline
\end{tabular}

\vspace{0.5em}
\begin{minipage}{0.96\linewidth}
\footnotesize
MAE denotes the mean absolute reconstruction error between the RL allocation
and the final M2 allocation. $\kappa_2(X)$ denotes the spectral condition
number of the reduced bin-level design matrix.
$^{\dagger}$The original mean ego-network resource in the final bin is
444.230. For bin assignment, resource values above 430 are capped at 430,
giving an effective mean of 426.088 for this bin.
\end{minipage}
\end{table}

\begin{table}[htbp]
\centering
\caption{Fitted M2 profiles for degree group $Q_2:d_i=3$.}
\label{tab:M2_Q2}
\small
\setlength{\tabcolsep}{3.5pt}
\begin{tabular}{ccccccc}
\hline
Resource bin &
Mean $\bar{R}_{\mathrm{ego}}$ &
$\omega_{\mathrm{Prop}}$ &
$\omega_{\mathrm{Equal}}$ &
$\omega_{\mathrm{Self}}$ &
MAE &
$\kappa_2(X)$ \\
\hline
$[2.687,67.799)$     & 27.249  & 0.9240 & 0.0651 & 0.0109 & 0.01635 & 1.9557 \\
$[67.799,266.366)$   & 192.990 & 0.9831 & 0.0140 & 0.0029 & 0.00981 & 2.0112 \\
$[266.366,284.069)$  & 274.182 & 1.0000 & 0.0000 & 0.0000 & 0.01035 & 1.8575 \\
$[284.069,315.028)$  & 298.200 & 1.0000 & 0.0000 & 0.0000 & 0.00939 & 2.1954 \\
$[315.028,317.950)$  & 316.300 & 1.0000 & 0.0000 & 0.0000 & 0.00828 & 2.3154 \\
$[317.950,323.091)$  & 319.827 & 1.0000 & 0.0000 & 0.0000 & 0.00982 & 2.5614 \\
$[323.091,340.222)$  & 332.544 & 1.0000 & 0.0000 & 0.0000 & 0.01089 & 2.9979 \\
$[340.222,365.412)$  & 354.069 & 1.0000 & 0.0000 & 0.0000 & 0.00847 & 3.2081 \\
$[365.412,376.102)$  & 372.534 & 1.0000 & 0.0000 & 0.0000 & 0.00549 & 2.7987 \\
$[376.102,385.386)$  & 382.640 & 1.0000 & 0.0000 & 0.0000 & 0.00495 & 1.8413 \\
$[385.386,405.036)$  & 399.902 & 0.9905 & 0.0000 & 0.0095 & 0.00567 & 2.9457 \\
$[405.036,414.046]$  & 411.598 & 0.9973 & 0.0000 & 0.0027 & 0.00676 & 3.2760 \\
\hline
\end{tabular}
\end{table}

\begin{table}[htbp]
\centering
\caption{Fitted M2 profiles for degree group
$Q_3:d_i=4$--$6$.}
\label{tab:M2_Q3}
\small
\setlength{\tabcolsep}{3.5pt}
\begin{tabular}{ccccccc}
\hline
Resource bin &
Mean $\bar{R}_{\mathrm{ego}}$ &
$\omega_{\mathrm{Prop}}$ &
$\omega_{\mathrm{Equal}}$ &
$\omega_{\mathrm{Self}}$ &
MAE &
$\kappa_2(X)$ \\
\hline
$[4.747,67.330)$     & 27.148  & 0.9401 & 0.0553 & 0.0046 & 0.01336 & 1.9527 \\
$[67.330,297.224)$   & 211.196 & 0.9776 & 0.0217 & 0.0007 & 0.00964 & 1.8976 \\
$[297.224,329.575)$  & 302.850 & 0.9990 & 0.0010 & 0.0000 & 0.00996 & 1.5752 \\
$[329.575,330.951)$  & 330.391 & 1.0000 & 0.0000 & 0.0000 & 0.00527 & 2.3301 \\
$[330.951,335.514)$  & 334.011 & 0.9998 & 0.0002 & 0.0000 & 0.00889 & 2.6181 \\
$[335.514,337.016)$  & 336.704 & 0.9584 & 0.0416 & 0.0000 & 0.00440 & 3.2923 \\
$[337.016,356.173)$  & 350.132 & 0.9859 & 0.0141 & 0.0000 & 0.00595 & 3.3384 \\
$[356.173,364.171)$  & 359.523 & 0.9894 & 0.0106 & 0.0000 & 0.00617 & 3.4767 \\
$[364.171,372.682)$  & 365.972 & 0.9714 & 0.0286 & 0.0000 & 0.00542 & 3.2897 \\
$[372.682,376.546)$  & 373.191 & 0.9547 & 0.0453 & 0.0000 & 0.00500 & 3.6555 \\
$[376.546,391.396)$  & 378.595 & 0.9991 & 0.0009 & 0.0000 & 0.00568 & 2.0373 \\
$[391.396,393.942]$  & 393.667 & 1.0000 & 0.0000 & 0.0000 & 0.00603 & 2.4938 \\
\hline
\end{tabular}
\end{table}

\begin{table}[htbp]
\centering
\caption{Fitted M2 profiles for degree group $Q_4:d_i=7$--$19$.}
\label{tab:M2_Q4}
\small
\setlength{\tabcolsep}{3.5pt}
\begin{tabular}{ccccccc}
\hline
Resource bin &
Mean $\bar{R}_{\mathrm{ego}}$ &
$\omega_{\mathrm{Prop}}$ &
$\omega_{\mathrm{Equal}}$ &
$\omega_{\mathrm{Self}}$ &
MAE &
$\kappa_2(X)$ \\
\hline
$[7.923,83.080)$     & 33.691  & 0.8711 & 0.1275 & 0.0014 & 0.00831 & 3.6776 \\
$[83.080,340.416)$   & 238.796 & 0.8859 & 0.1135 & 0.0006 & 0.00579 & 4.4471 \\
$[340.416,340.469)$  & 340.436 & 0.9324 & 0.0676 & 0.0000 & 0.00516 & 3.9318 \\
$[340.469,340.642)$  & 340.602 & 0.9002 & 0.0998 & 0.0000 & 0.00702 & 3.1233 \\
$[340.642,349.439)$  & 342.084 & 0.9084 & 0.0916 & 0.0000 & 0.00666 & 3.2873 \\
$[349.439,357.178)$  & 356.796 & 0.9270 & 0.0730 & 0.0000 & 0.00500 & 4.3164 \\
$[357.178,385.594)$  & 363.028 & 0.9046 & 0.0954 & 0.0000 & 0.00627 & 4.6127 \\
$[385.594,387.393)$  & 387.280 & 0.6829 & 0.3171 & 0.0000 & 0.01160 & 6.4937 \\
$[387.393,406.338)$  & 394.225 & 0.8030 & 0.1961 & 0.0010 & 0.01072 & 7.0853 \\
$[406.338,406.728)$  & 406.665 & 0.9749 & 0.0224 & 0.0027 & 0.00816 & 8.1860 \\
$[406.728,407.539)$  & 407.312 & 0.8020 & 0.1972 & 0.0008 & 0.00804 & 8.9819 \\
$[407.539,421.490]$  & 409.509 & 0.7453 & 0.2547 & 0.0000 & 0.00755 & 9.5539 \\
\hline
\end{tabular}
\end{table}

\section{Analysis of Steady State}
\label{Appendix:Analysis of Steady State}
\setcounter{equation}{0}
\renewcommand{\theequation}{C.\arabic{equation}}

In this section, we provide two analytical benchmarks for the model.
We first derive an upper bound for the steady-state resource level under a homogeneous approximation, and then analyze the degree-induced Gini coefficient on BA scale-free networks under full cooperation.

\subsection{Steady-state resource analysis}

To obtain an analytical benchmark for the resource level under persistent cooperation, we consider a homogeneous setting in which the interaction network is $d$-regular and all individuals remain cooperative at steady state. In addition, we assume that resource allocation within each local pool is uniform (Equal baseline). Under this approximation, all individuals are statistically equivalent, and we denote their common resource level at time $t$ by $R(t)$.

Since every node has degree $d$, each individual participates in exactly $d+1$ local pools, including its own pool and those centered at its neighbours. Under full cooperation, the contribution of each individual is
\begin{equation}
e(t)=d+1+\alpha\bigl(R(t)-(d+1)\bigr).
\end{equation}
Because the network is homogeneous, each pool receives the same total contribution,
\begin{equation}
P(t)=\sum_{j \in N_i \cup \{i\}} \frac{e_j(t)}{d_j+1}=e(t).
\end{equation}

The allocable resource in each pool is therefore
\begin{equation}
G(t)=\min\bigl\{(1+r)e(t),\,p_c(d+1)\bigr\}.
\end{equation}
Under uniform allocation, each pool returns $G(t)/(d+1)$ to each of the $d+1$ members in its local group. Since each individual belongs to exactly $d+1$ pools, the total received resource is
\begin{equation}
u(t)=G(t).
\end{equation}

Moreover, under the all-cooperation steady state we must have $R(t)\ge d+1>d$, and thus the consumption function lies in the linear regime:
\begin{equation}
c(t)=d+\tau\bigl(R(t)-d\bigr).
\end{equation}
Hence, the resource update equation becomes
\begin{equation}
R(t+1)=R(t)+u(t)-e(t)-c(t).
\end{equation}

At steady state, $R(t+1)=R(t)=R^*$, so
\begin{equation}
e^*+c^*=u^*=G^*.
\end{equation}
Since
\begin{equation}
G^*=\min\bigl\{(1+r)e^*,\,p_c(d+1)\bigr\}\le p_c(d+1),
\end{equation}
it follows that
\begin{equation}
e^*+c^*\le p_c(d+1).
\end{equation}
Substituting the expressions for $e^*$ and $c^*$ yields
\begin{equation}
d+1+\alpha\bigl(R^*-(d+1)\bigr)+d+\tau\bigl(R^*-d\bigr)\le p_c(d+1).
\end{equation}
After rearranging, we obtain the upper bound
\begin{equation}
R^*
\le
\frac{(\tau-1)d+(p_c+\alpha-1)(d+1)}{\alpha+\tau}.
\label{eq:steady_bound}
\end{equation}

Therefore, under homogeneous averaging and persistent cooperation, the steady-state resource level of an individual is bounded above by Eq.~\eqref{eq:steady_bound}. Equality holds if and only if the local pool operates on the saturated branch,
\begin{equation}
(1+r)e^* \ge p_c(d+1),
\end{equation}
so that
\begin{equation}
G^*=p_c(d+1).
\end{equation}

Under the default parameter setting
$N=50$, $r=0.5$, $T=2\times10^4$, $R_0=20$, $p_c=50$, $\alpha=0.5$, $\tau=0.05$, and $\kappa=1$,
and using the homogeneous benchmark degree $d=\langle k\rangle=4$,
Eq.~\eqref{eq:steady_bound} yields
\begin{equation}
R^*
=
\frac{(\tau-1)d+(p_c+\alpha-1)(d+1)}{\alpha+\tau}
=
\frac{(0.05-1)\cdot 4+(50+0.5-1)\cdot 5}{0.5+0.05}
\approx 443.09.
\end{equation}
Moreover, the corresponding allocable resource per local pool is
\begin{equation}
G^*=p_c(d+1)=50\times 5=250.
\end{equation}
Since
\begin{equation}
(1+r)e^*\approx 336.07 > 250,
\end{equation}
the steady state indeed lies on the saturated branch, which justifies the equality case in the above derivation.

This result provides a homogeneous analytical benchmark for the steady-state resource level. We next extend the analysis to heterogeneous BA scale-free networks, where degree heterogeneity induces nontrivial resource inequality even under full cooperation.
\subsection{Steady-state scale-free Gini analysis}
We next analyze the resource inequality induced by degree heterogeneity under the all-cooperation steady state on a Barab\'asi--Albert (BA) scale-free network. We assume that resources in each local pool are distributed equally among the members of the corresponding local group, and that the accumulated resource is at steady state.

Let individual $i$ have degree $d_i$, and define
\begin{equation}
H_i := d_i + 1.
\end{equation}
Then $H_i$ is the size of the local group associated with individual $i$, including itself and its neighbours.

Under full cooperation, individual $i$ contributes
\begin{equation}
e_i(t)=d_i+1+\alpha\bigl(R_i(t)-(d_i+1)\bigr).
\end{equation}
Moreover, in the high-resource regime $R_i(t)>d_i$, the consumption of individual $i$ is
\begin{equation}
c_i(t)=d_i+\tau\bigl(R_i(t)-d_i\bigr).
\end{equation}

For pool $j$, the capacity upper bound is
\begin{equation}
P_{\max,j}(t)=p_c\,N_{c,j}(t)\,\rho_{c,j}(t).
\end{equation}
Under full cooperation, we have $N_{c,j}(t)=d_j+1=H_j$ and $\rho_{c,j}(t)=1$, so that
\begin{equation}
P_{\max,j}(t)=p_cH_j.
\end{equation}
Hence, on the saturated branch,
\begin{equation}
G_j^*=p_cH_j.
\end{equation}

Since pool $j$ distributes its allocable resource equally among the $H_j$ members of its local group, each member receives
\begin{equation}
\frac{G_j^*}{H_j}=p_c
\end{equation}
from that pool. Individual $i$ belongs to exactly $H_i=d_i+1$ pools, including its own pool and those of its neighbours. Therefore, its total received resource at steady state is
\begin{equation}
u_i^*=p_cH_i=p_c(d_i+1).
\end{equation}

At steady state, $R_i(t+1)=R_i(t)=R_i^*$, and thus
\begin{equation}
u_i^*=e_i^*+c_i^*.
\end{equation}
Substituting the above expressions gives
\begin{equation}
p_c(d_i+1)
=
d_i+1+\alpha\bigl(R_i^*-(d_i+1)\bigr)
+
d_i+\tau\bigl(R_i^*-d_i\bigr).
\end{equation}
After rearranging, we obtain
\begin{equation}
R_i^*
=
\frac{(\tau-1)d_i+(p_c+\alpha-1)(d_i+1)}{\alpha+\tau}.
\label{eq:Ri_star_sf}
\end{equation}

Equation~\eqref{eq:Ri_star_sf} shows that the steady-state resource is an affine function of degree. Equivalently, by writing $H_i=d_i+1$, we obtain
\begin{equation}
R_i^*=aH_i+b,
\end{equation}
where
\begin{equation}
a=\frac{p_c+\alpha+\tau-2}{\alpha+\tau},
\qquad
b=\frac{1-\tau}{\alpha+\tau}.
\label{eq:affine_R}
\end{equation}

We now consider a BA scale-free network with attachment parameter $m=2$. In the large-network limit, the degree distribution is
\begin{equation}
\mathbb{P}(d=k)=\frac{12}{k(k+1)(k+2)},
\qquad k\ge 2.
\end{equation}
Hence, for $H=d+1$,
\begin{equation}
\mathbb{P}(H=n)=\frac{12}{(n-1)n(n+1)},
\qquad n\ge 3.
\end{equation}
For this distribution, one has
\begin{equation}
\mathbb{E}[H]=5,
\end{equation}
and the corresponding Gini coefficient is
\begin{equation}
G(H)=24-\frac{12}{5}\pi^2\approx 0.31295.
\label{eq:Gini_H}
\end{equation}

Since $R_i^*$ is an affine transform of $H_i$, the Gini coefficient of the steady-state resource distribution satisfies
\begin{equation}
G(R^*)
=
\frac{a\,\mathbb{E}[H]}{a\,\mathbb{E}[H]+b}\,G(H).
\end{equation}
Substituting Eqs.~\eqref{eq:affine_R} and \eqref{eq:Gini_H}, we obtain
\begin{equation}
G(R^*)
=
\frac{5(p_c+\alpha+\tau-2)}
{5(p_c+\alpha+\tau-2)+(1-\tau)}
\left(24-\frac{12}{5}\pi^2\right).
\label{eq:Gini_R_star}
\end{equation}

Under the default parameter setting
$N=50$, $r=0.5$, $T=2\times10^4$, $R_0=20$, $p_c=50$, $\alpha=0.5$, $\tau=0.05$, and $\kappa=1$,
Eq.~\eqref{eq:Gini_R_star} gives
\begin{equation}
G_{BA}(R^*)
=
\frac{5(50+0.5+0.05-2)}
{5(50+0.5+0.05-2)+(1-0.05)}
\left(24-\frac{12}{5}\pi^2\right)
\approx 0.3117.
\end{equation}
This value is derived from the asymptotic degree distribution of the BA network with $m=2$, and therefore serves as a large-network theoretical benchmark. For a finite network such as $N=50$, the observed Gini coefficient may deviate slightly from this value.

Therefore, under equal baseline and the all-cooperation steady state, the asymptotic theoretical Gini coefficient on a BA scale-free network with $m=2$ is approximately $0.312$.

\begin{figure}
    \centering
    \includegraphics[width=0.9\linewidth]{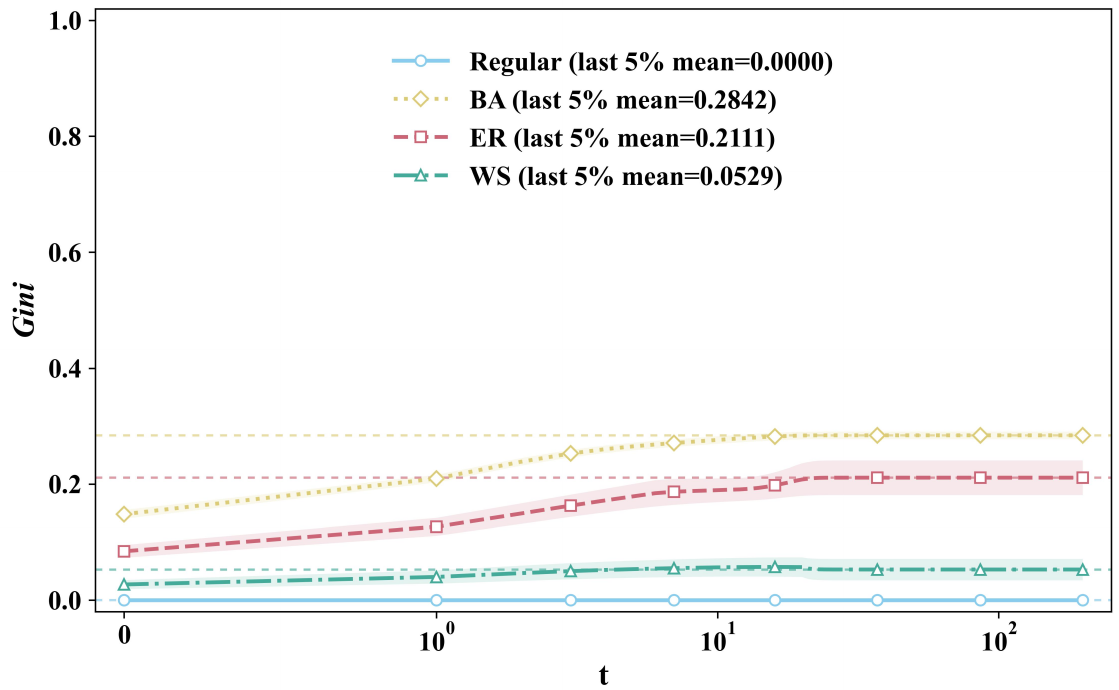}
    \caption{Simulated resource-inequality dynamics under persistent cooperation and equal allocation across network topologies on the same 50 graph instances
    used in the Fig.~\ref{fig:five_policy_diagram_7} evaluation.}
    \label{fig:Appendix_Gini}
\end{figure}
\subsection{Extension to ER and WS networks}
The above BA analysis can be extended to other network topologies without repeating the full steady-state derivation. 
The key observation is that, under full cooperation, equal local redistribution, and the saturated-pool regime, the steady-state resource level is an affine function of the local group size $H_i=d_i+1$, namely
\begin{equation}
R_i^*=aH_i+b,
\end{equation}
where
\begin{equation}
a=\frac{p_c+\alpha+\tau-2}{\alpha+\tau},
\qquad
b=\frac{1-\tau}{\alpha+\tau}.
\end{equation}
Therefore, the ideal all-cooperation Gini coefficient can be written as
\begin{equation}
G(R^*)=
\frac{a\bar H}{a\bar H+b}G(H),
\end{equation}
where $\bar H=N^{-1}\sum_i H_i$ and $G(H)$ denotes the Gini coefficient of the local group size distribution.

For the ER network used in our simulations, we set $N=50$ and $p_{\mathrm{ER}}=4/(50-1)$. 
Thus, the expected degree is $\langle d\rangle=4$, and the degree distribution is approximated by
\begin{equation}
d\sim \mathrm{Binomial}\left(49,\frac{4}{49}\right).
\end{equation}
Substituting this distribution into the above formula gives
\begin{equation}
G_{\mathrm{ER}}(R^*)\approx 0.2122.
\end{equation}

For the WS network, we set $N=50$, $k=4$, and rewiring probability $q=0.1$. 
Since the WS degree distribution depends on the random rewiring realization, we use the corresponding expected degree-heterogeneity benchmark. 
This gives
\begin{equation}
G_{\mathrm{WS}}(R^*)\approx 0.0564.
\end{equation}

Given that a finite-N truncation of the tail might reduce the Gini index appreciably, we conducted simulations to obtain more precise results as a simulated benchmark in Fig.~\ref{fig:Appendix_Gini}.
All individuals were constrained to remain cooperative, and each pool distributed its resources uniformly among the focal node and its neighbors. The curves show the mean resource Gini coefficient over 50 simulations, with shaded bands denoting ±1 standard deviation. Horizontal reference lines indicate the steady-state Gini level, calculated by averaging the ensemble-mean trajectory over the final $5\%$ of the 200 simulation steps. Regular, BA, ER, and WS denote random regular, Barabási–Albert, Erdős–Rényi, and Watts–Strogatz networks, respectively, with \(N=50\). The time axis is displayed on a symmetric logarithmic scale.

These values show that the ideal all-cooperation Gini coefficient is mainly determined by the degree heterogeneity of the underlying topology. Even under persistent cooperation and equal local redistribution, global resource equality is not guaranteed in heterogeneous networks, because high-degree nodes participate in more local pools and therefore receive resources from more sources. Consequently, BA networks induce the largest degree-driven inequality, ER networks induce a moderate level of inequality, whereas WS networks with low rewiring probability remain close to the regular-network benchmark.

% \noindent\textit{Notes.}
% (1) The actor global encoder and the critic state encoder use the same input format but do not share parameters.
% (2) The ego-network allocation head is applied repeatedly to every focal node with shared weights.
% (3) The twin critics share the same architecture but are independently parameterized.
\section{Interpolation baseline} 
\label{Appendix:Interpolation baseline}
\setcounter{equation}{0}
\renewcommand{\theequation}{D.\arabic{equation}}

To provide an additional reference \cite{koster2025deep} for evaluating the learned allocation mechanism, we consider an Interpolation baseline that adaptively combines the Equal and Proportional allocation mechanisms according to the current resource level of each local pool. Specifically, for resource pool $i$ at time $t$, the allocation vector generated by the Interpolation mechanism is defined as
\begin{equation}
A_i^{\mathrm{Interp}}(t)
=
w_i(t) A_i^{\mathrm{Equal}}(t)
+
\left[1-w_i(t)\right] A_i^{\mathrm{Prop}}(t),
\end{equation}
where $A_i^{\mathrm{Equal}}(t)$ and $A_i^{\mathrm{Prop}}(t)$ denote the allocation vectors generated by the Equal and Proportional mechanisms, respectively. The interpolation weight $0\leq w_i(t)\leq1$ is determined by the current allocable resources of the pool as
\begin{equation}
w_i(t)
= \min\{1,
\left(
\frac{G_i(t)}
{\langle k+1\rangle p_c}
\right)^{k_I}\},
\end{equation}
where $G_i(t)$ denotes the current allocable resources of pool $i$, $\langle k\rangle$ is the average degree of the network, and $p_c$ is the pool-capacity coefficient. The exponent $k_I$ controls the sensitivity of the interpolation weight to the current pool-resource level.

The exponent $k_I$ is a fixed hyperparameter rather than a parameter learned by the reinforcement learning agent. We empirically use $k_I=13$ for all experiments involving the Interpolation mechanism.
The same Interpolation rule is included as one component of the heuristic teacher set during GNN-RL training. In the additional baseline evaluation, however, it is directly applied as a standalone allocation mechanism without involving the learned GNN-RL policy. All other network settings, evolutionary dynamics, initial conditions, and evaluation procedures are kept identical to those used for the other allocation mechanisms. 

The Interpolation mechanism is used both as a standalone baseline and as one component of the heuristic teacher set during training. The standalone evaluation allows us to separate the performance of this prior heuristic from that of the learned RL policy and the subsequently distilled M1 and M2 mechanisms. As discussed in Sections 4.4 and 5, the Interpolation teacher provides useful early-stage training guidance, whereas the final learned policy exhibits additional resource- and structure-dependent allocation patterns that are not encoded by this baseline.

% This comparison reveals that the Interpolation teacher, as a crucial meta-heuristic teacher, plays a significant role in the training process. However, the RL-Agent's strategy, discovered through learning based on heterogeneous network relationships, outperforms that of the Interpolation teacher. Furthermore, our strategy distillation results for M1 and M2 reveal a strategy that differs significantly from that of the Interpolation teacher.

\section{Additional Generalization Tests}
\label{Appendix:Additional Generalization Tests}
\setcounter{equation}{0}
\renewcommand{\theequation}{E.\arabic{equation}}

To further examine the robustness of the learned and interpretable allocation mechanisms, we conduct additional generalization tests by varying key environmental parameters. Specifically, we change the pool-capacity coefficient $p_c$, the Fermi stochasticity parameter $\kappa$, the network size $N$, and the pool growth factor $r$, while keeping the remaining settings unchanged. These tests evaluate whether the qualitative advantage of M2 persists beyond the default parameter configuration used in the main text.
It should be noted that changing $p_c$ directly affects the scale of the average accumulated resource $\bar{R}(t)$, because $p_c$ controls the upper capacity of local resource pools. Therefore, to make the results comparable across different parameter settings, we report the normalized average accumulated resource defined as
\begin{equation}
\bar{R}_{\mathrm{norm}}(t)=\frac{\bar{R}(t)}{R^*}, \notag
\end{equation}
where $R^*$ denotes the corresponding theoretical benchmark resource level under the given parameter setting in~\ref{Appendix:Analysis of Steady State}. In addition, the resource-dependent binning used in M1 and M2 is also performed after this normalization. That is, the normalized resource level is first mapped into the corresponding resource bin, and then the allocation weights of the mixture mechanism are determined accordingly. This treatment ensures that the mixture mechanisms are compared under a unified resource scale when the environmental capacity changes.

\begin{figure}[p]
    \centering
    \includegraphics[width=0.95\linewidth]{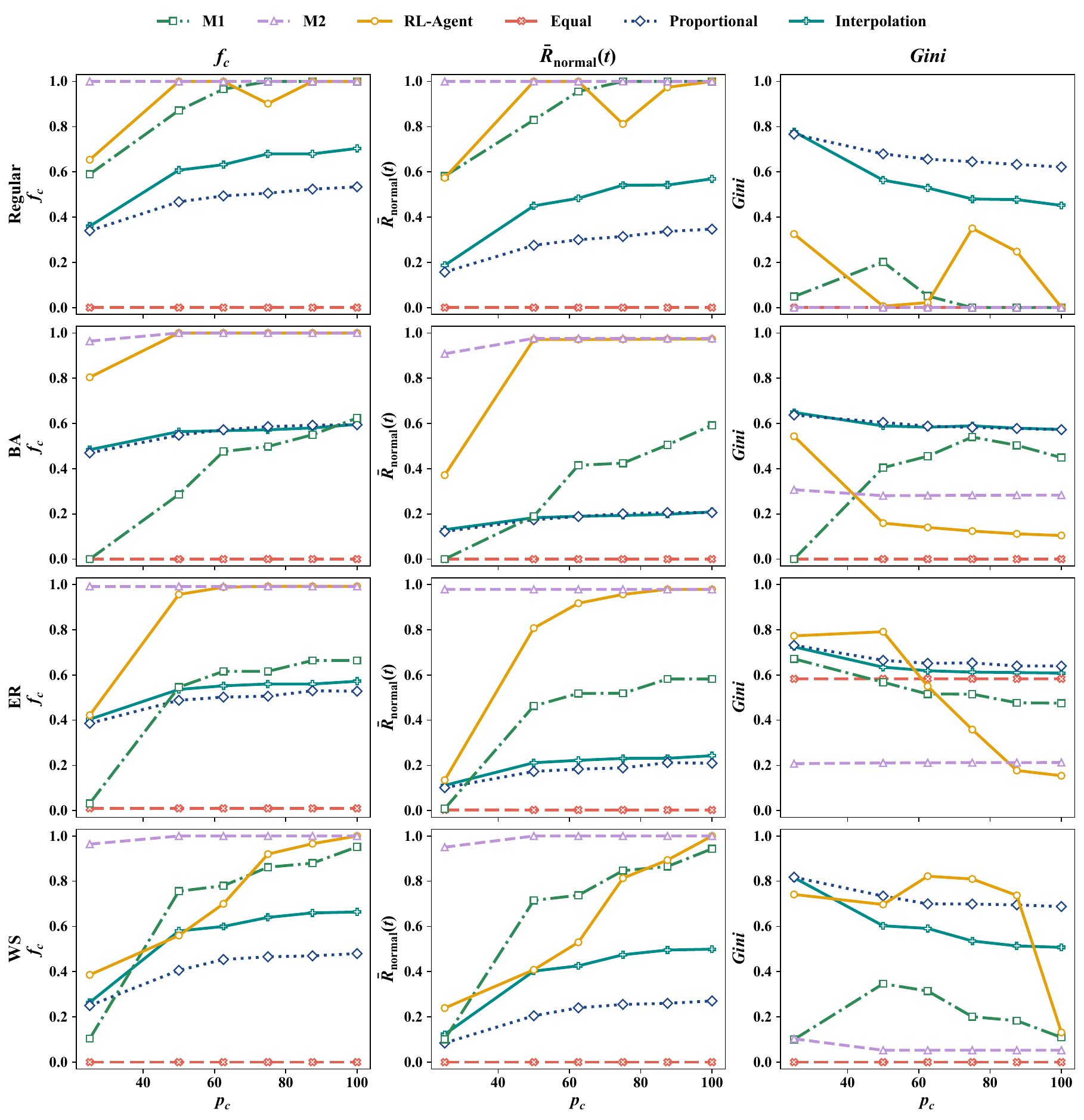}
    \caption{Generalization test under different values of the pool-capacity coefficient $p_c$. The normalized average accumulated resource $\bar{R}_{\mathrm{norm}}(t)$ is used to make results comparable across different resource-capacity scales. M2 maintains stable performance across different $p_c$ values, whereas the original RL-Agent exhibits larger fluctuations.}
    \label{fig:generalization_pc}
\end{figure}

\begin{figure}[p]
    \centering
    \includegraphics[width=0.95\linewidth]{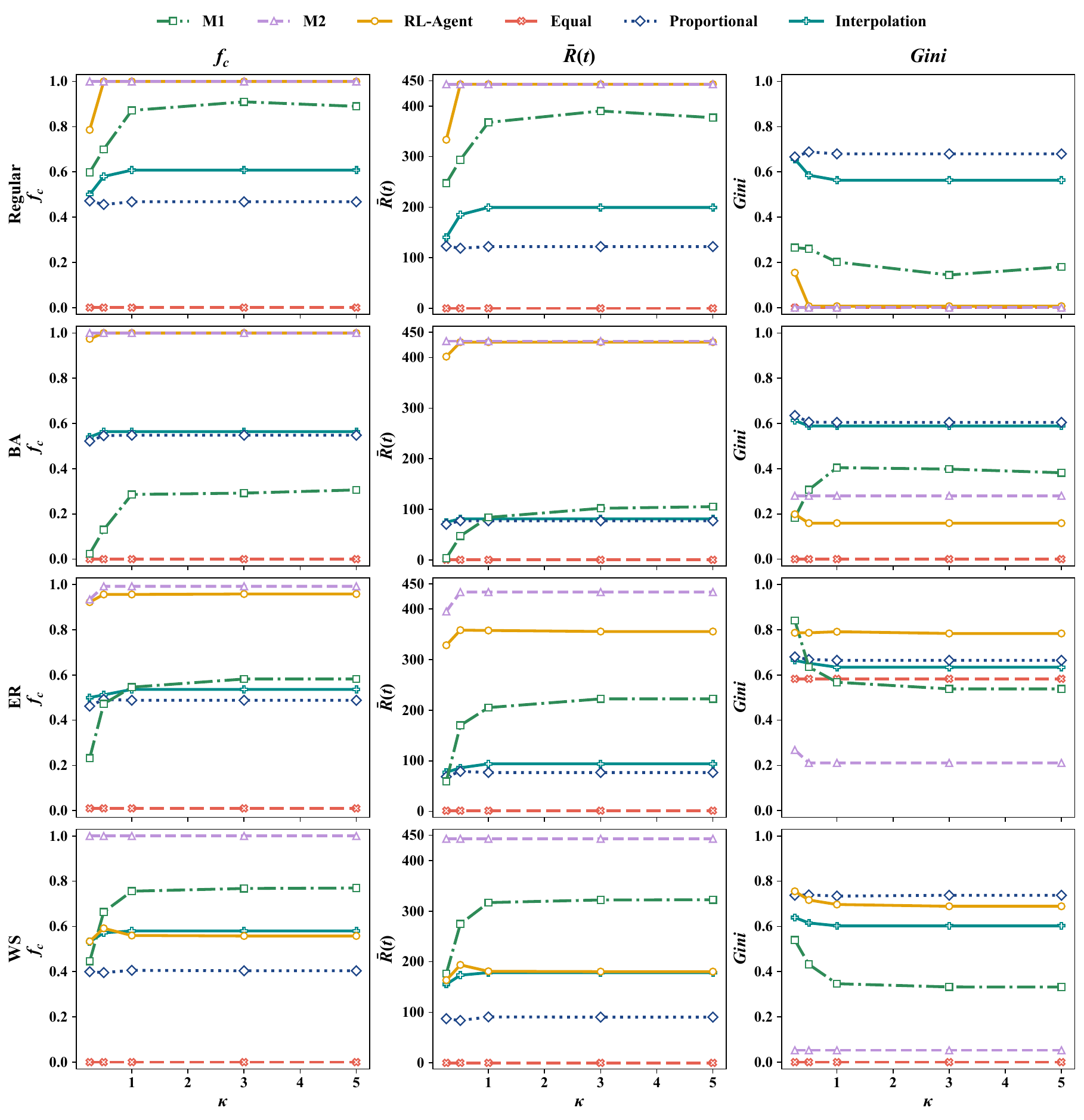}
    \caption{Generalization test under different values of the Fermi stochasticity parameter $\kappa$. M2 remains robust when the stochasticity of strategy updating changes, while the original RL-Agent shows more visible sensitivity to the parameter shift.}
    \label{fig:generalization_kappa}
\end{figure}

\begin{figure}[p]
    \centering
    \includegraphics[width=0.95\linewidth]{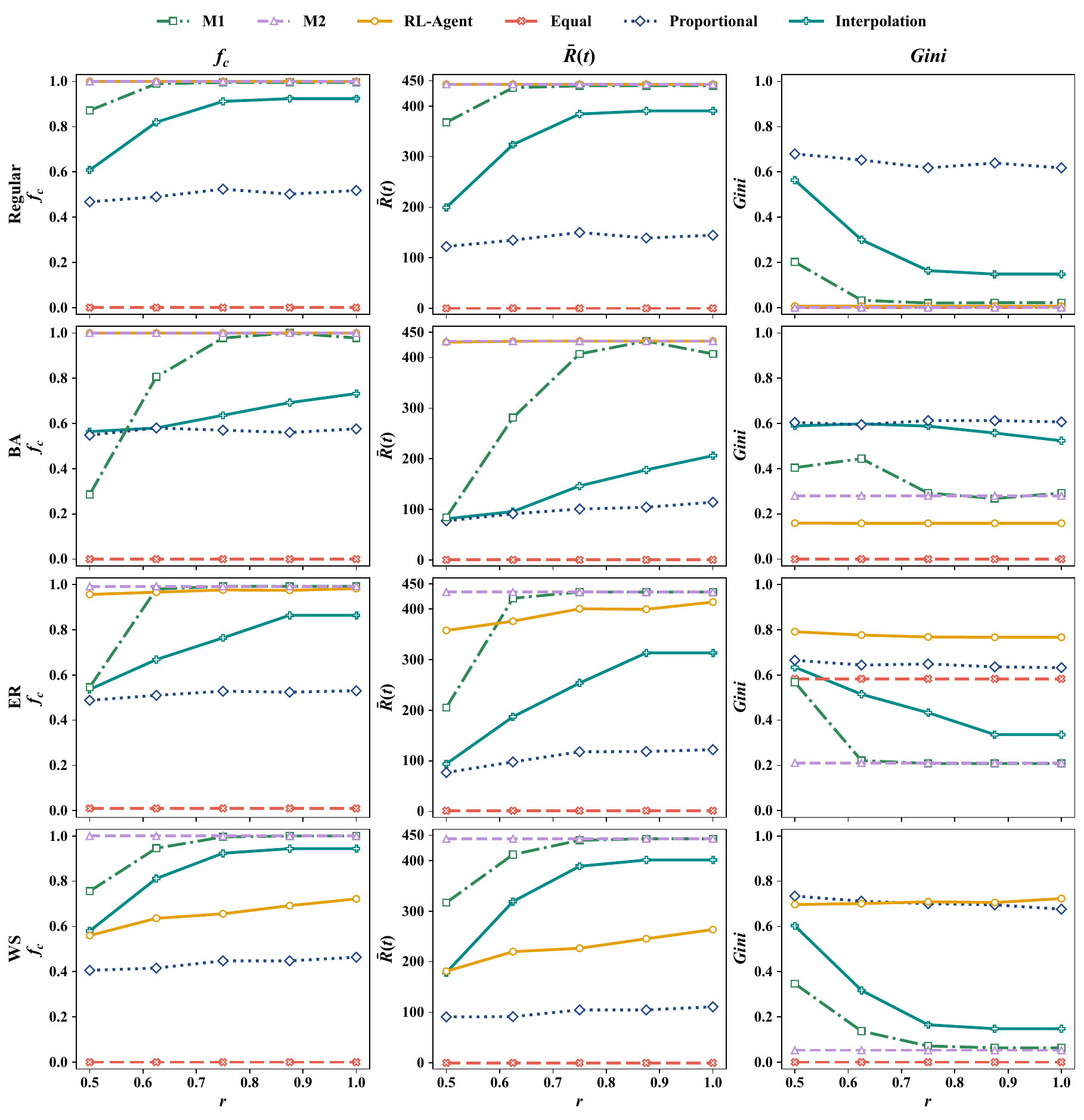}
    \caption{Generalization test under different values of the pool growth factor $r$. M2 preserves its qualitative advantage across different resource-growth intensities, indicating that the extracted degree-conditioned mechanism generalizes beyond the default growth setting.}
    \label{fig:generalization_r}
\end{figure}

\begin{figure}[p]
    \centering
    \includegraphics[width=0.95\linewidth]{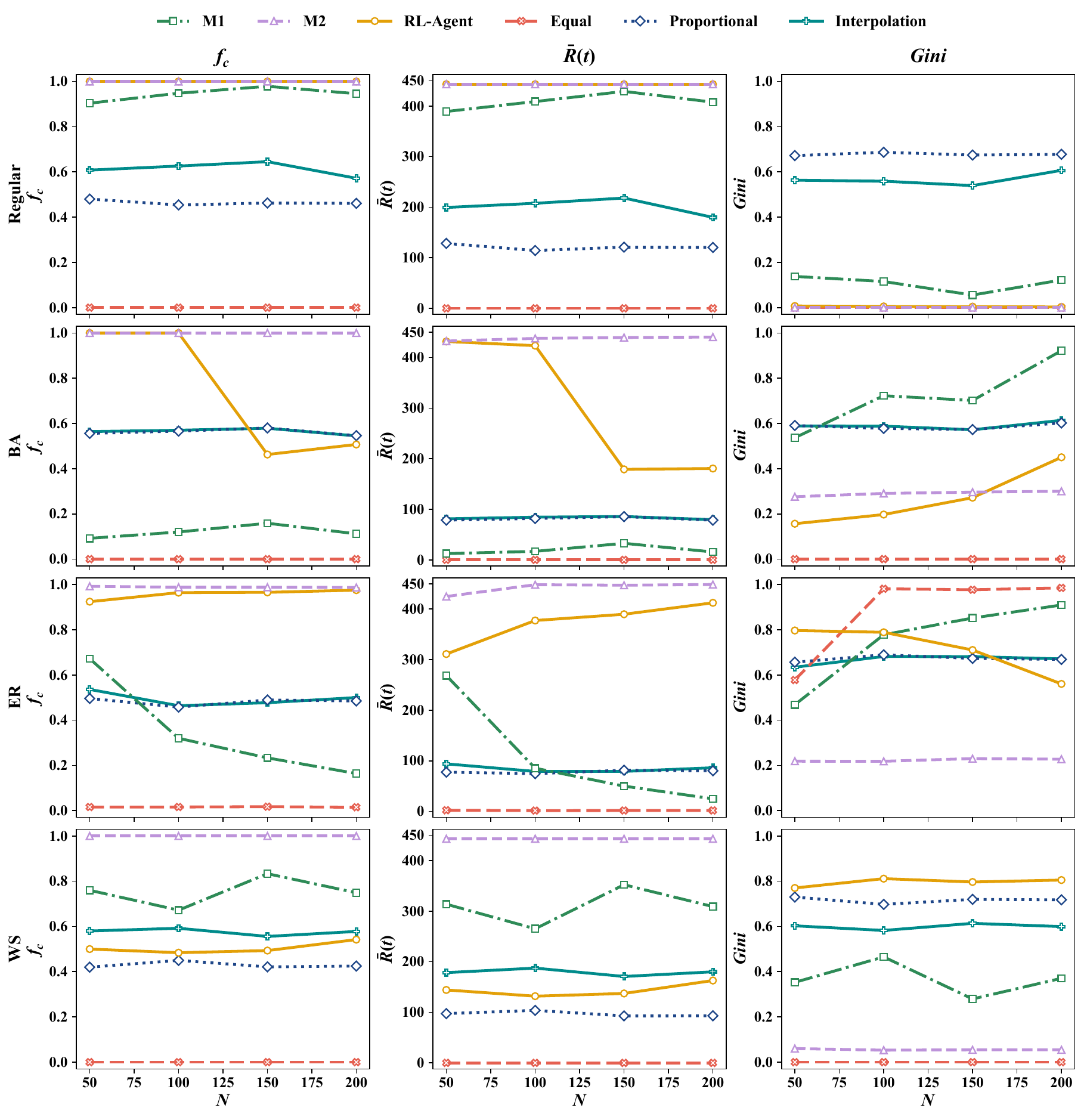}
    \caption{Generalization test under different values of the network sizes $N$. M2 preserves its qualitative advantage across different network sizes, indicating that the extracted degree-conditioned mechanism generalizes beyond the default network size setting.}
    \label{fig:generalization_n}
\end{figure}

Figures~\ref{fig:generalization_pc}--\ref{fig:generalization_n} report the generalization results under variations of $p_c$, $\kappa$, $r$ and $N$, respectively. Overall, M2 exhibits stable performance across different parameter values. Its cooperation level, average accumulated resource, and Gini coefficient remain largely unchanged under these environmental shifts. By contrast, the original RL-Agent shows more visible performance fluctuations when the parameters deviate from the default setting. These results suggest that M2 does not merely overfit the default simulation environment, but captures a more robust allocation principle that generalizes across changes in resource capacity, stochasticity of strategy update, and pool growth intensity.

% \begin{figure}[ht]
%     \centering
%     \includegraphics[width=0.95\linewidth]{Gerneraliztion_pc.pdf}
%     \caption{Caption}
%     \label{fig:enter-label}
% \end{figure}

% \begin{figure}[ht]
%     \centering
%     \includegraphics[width=0.95\linewidth]{Gerneraliztion_kappa.pdf}
%     \caption{Caption}
%     \label{fig:enter-label}
% \end{figure}

% \begin{figure}[ht]
%     \centering
%     \includegraphics[width=0.95\linewidth]{Gerneraliztion_r.pdf}
%     \caption{Caption}
%     \label{fig:enter-label}
% \end{figure}

\clearpage
\bibliographystyle{apalike}
\bibliography{refs}
\end{document}